\documentclass[usenatbib,twocolumn]{mnras}
\usepackage{graphicx,amsmath,amssymb,enumitem,verbatim,sidecap}
\begin{document}

\title[Clusters in the Disperse cosmic web]{Clusters in the {\sc Disperse} cosmic web}

\author[Clusters in the Disperse cosmic web]{J.D. Cohn${}^{1,2}$\thanks{E-mail: jcohn@berkeley.edu}\\
${}^1$Space Sciences Laboratory 
  University of California, Berkeley, CA 94720, USA\\
${}^2$Theoretical Astrophysics Center,
  University of California, Berkeley, CA 94720, USA}
\maketitle
\begin{abstract} 
  Galaxy cluster mass halos (``clusters'') in a dark matter simulation
  are matched to nodes in several different cosmic webs found using the {\sc Disperse} cosmic web finder. The webs have different simulation smoothings and {\sc Disperse} parameter choices; for each, 4 methods are considered for matching {\sc
    Disperse} nodes to clusters.  For most of the webs, 
 {\sc Disperse} nodes outnumber clusters, but not every
  cluster has a {\sc Disperse} node match (and sometimes $>1$ cluster matches to the same {\sc Disperse} node). The clusters frequently lacking a matching {\sc
    Disperse} node have a different distribution of local shear trends
  and perhaps merger histories.   It might
  be interesting to see in what other ways, e.g., observational
  properties, these clusters differ.  For the webs 
  with smoothing $\leq$ 2.5 $Mpc/h$, and all but the most restrictive matching criterion, $\sim$3/4 of
  the clusters always have a {\sc Disperse} node counterpart. 
  The nearest cluster to a given {\sc
    Disperse} node and vice versa, within twice
  the smoothing length, obey a cluster mass-{\sc Disperse}
  node density relation.  Cluster pairs where both clusters match {\sc Disperse}
  nodes can also be assigned the filaments between those nodes, but as the web and matching methods are varied, most such filaments do not remain.
  There is an enhancement of subhalo counts and halo mass between cluster pairs,
  averaging over cluster pairs assigned {\sc Disperse} filaments increases the enhancement.   The approach here also lends
  itself to comparing nodes across many cosmic web constructions,
  using the fixed underlying cluster distribution to make a
  correspondence.
\end{abstract}
  
The cosmic web \citep{BonKofPog96,BonMye96,PogBonKof98} exists within
many tracers of large scale structure in the universe, from galaxies
to dark matter to gas, and underpins the evolution of large scale
structure. In the many years since its discovery, numerous ways of
characterizing the web and identifying its components (nodes,
filaments, voids and walls) have been proposed. Several of these web
finders are compared in \citet{Lib18}, earlier comparisons include,
e.g., \citet{Lec16}; see also the recent meetings
\citep{van16,Higgs19}.

In large scale structure, galaxy cluster mass halos (``clusters'' henceforth) are
the most massive bound structures; often clusters are referred to as
nodes in the cosmic web.  This correspondence frequently holds: in the web finder comparison of \citet{Lib18}, most of the
clusters lie in cosmic web nodes in all definitions
considered. Properties of the
relation between clusters and different variants of the web have been studied for a long
time, for instance in the context of the peak-patch algorithm
\citep{BonMye96}, or using the Gaussian field methods of \citet{BBKS},
also see the review of \citet{vanBon08}.  Clusters have also been used
to infer initial conditions, which are then run forward in time to get
large scale properties including the cosmic web
\citep[e.g.,][]{Bos16,Bosth16}.

In this note, cluster-node correspondences and properties of filaments linking
clusters are explored in more detail, for nodes and filaments in cosmic webs
found via {\sc Disperse} within a single dark matter simulation.
The aim here is to see how nodes in these cosmic webs
correspond to clusters, and from there, how the cluster pairs correspond to
node pairs which are connected by filaments.  The initial expectation was that
the clusters would pick out special nodes in the cosmic web(s) and that the filaments would pick out special associated cluster pairs.  However, for the webs
and cluster-node matchings used here, not all clusters had a node
counterpart. So in addition to clusters picking out certain nodes in
the webs, the webs picked out certain clusters.

The underlying dark matter simulation and the
web finder {\sc Disperse} are described in \S\ref{sec:simfinder}, as
well as a Hessian based node finder (used later as part of one
matching method). In \S\ref{sec:associate} a few methods to match
clusters and {\sc Disperse} nodes are suggested, applied and compared.
Cluster pairs are matched to {\sc Disperse} filaments in
\S\ref{sec:clusclusfil} and properties of cluster pairs with and without these filaments are
explored in \S\ref{sec:cluspair}. \S\ref{sec:disc} concludes.

\section{Simulation data and web finders}
\label{sec:simfinder}
\subsection{Simulation database}
The dark matter density map and halo and subhalo samples are taken
from the publicly available Millennium simulation \citep{Spr05} and
its database \citep{LemSpr06}.  The simulation corresponds to a fixed
time box with side 500 $Mpc/h$ and $2160^3$ particles; redshift
$z=0.12$ (step 58) is used.  The densities (relative to mean) are
available in the simulation on a $256^3$ pixel grid, with Gaussian
smoothing radii of 1.25 $Mpc/h$, 2.5 $Mpc/h$, 5 $Mpc/h$.\footnote{For
  1.25 Mpc/h smoothing, 
 `` select a.phkey, a.g1\_25 from MField..MField as a where a.snapnum=58
  order by a.phkey''.  For larger smoothings, substitute $g2\_5$ and
  $g\_5$.}.  The commonly used 2 $Mpc/h$ smoothing
\citep[e.g.,][]{Hah07a,Hah07b} is included by further smoothing the
1.25 $Mpc/h$ overdensities.  The halos are found via Friends of
Friends (FoF) \citep{DEFW} with linking length 0.2, and the subhalos are found via {\sc SUBFIND} \citep{Spr05}.
Subhalos (considered in \S 4.2, \S 4.3) can be
central, satellite or orphans; their 
virial mass is the mass of the FoF group they were in when they were
last a central subhalo. The 2898 massive halos with $M_{\rm vir} \geq
10^{14} M_\odot$ are taken to be clusters\footnote{Cluster mass halos are
  downloaded via ``select a.galaxyId,
  a.phKey, a.mvir, a.x, a.y, a.z, a.type from
  MPAGalaxies..deLucia2006a as a where a.phKey between NNN and MMM and
  a.mvir$>$=7300 and a.type=0 and a.snapnum=58'', where NNN, MMM
  select a phKey subset, and  mvir is virial
  mass of the FoF subhalo. This is done for all phKeys by doing for
  several ranges of NNN, MMM.}.  The Planck
cosmology \citep{Pla18} Millennium simulation is created from the
original Millennium simulation by shifts and scalings
\citep{AngWhi10,AngHil15}, however, the rescaled density field grid is
unavailable.  So the cosmological parameters are unfortunately
outdated ($\Omega_m$ = 0.25, $\Omega_b$ = 0.045, h = 0.73, n=1,
$\sigma_8$ = 0.9).  However, the particular cosmology is likely irrelevant to
the qualitative questions of interest here.
\begin{figure*}
\begin{center}
\resizebox{3.2in}{!}{\includegraphics{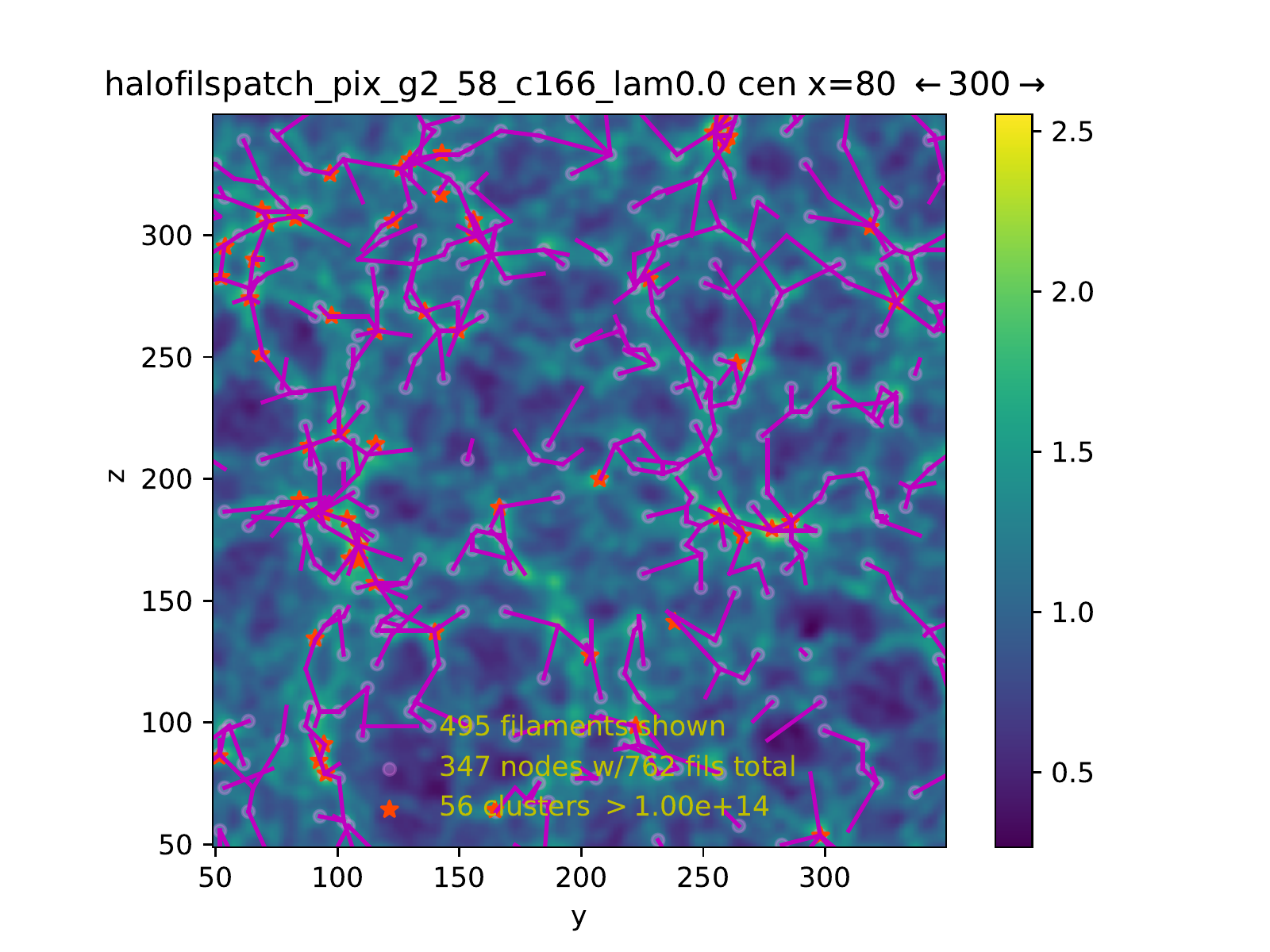}}
\resizebox{3.2in}{!}{\includegraphics{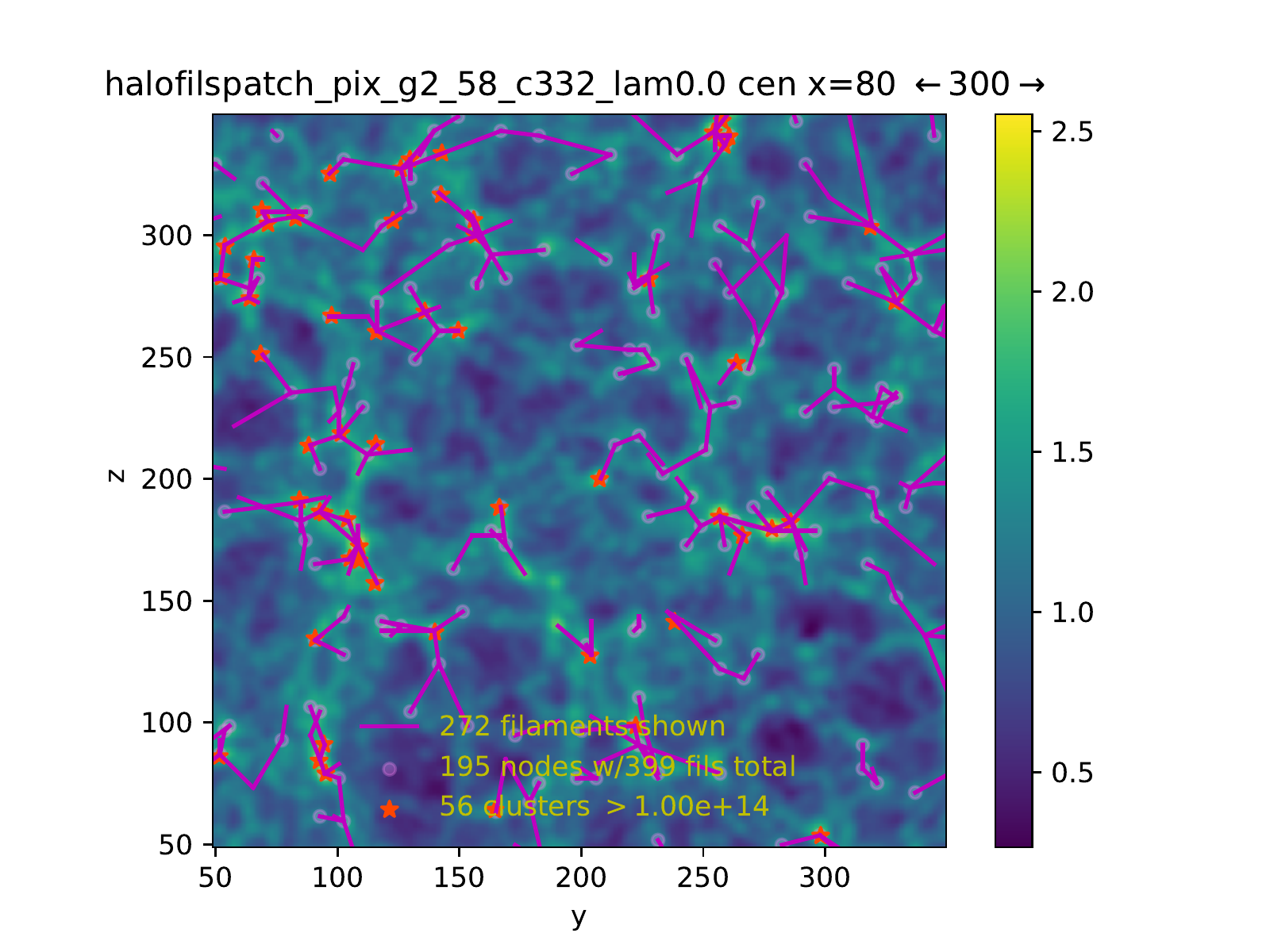}}
\resizebox{3.2in}{!}{\includegraphics{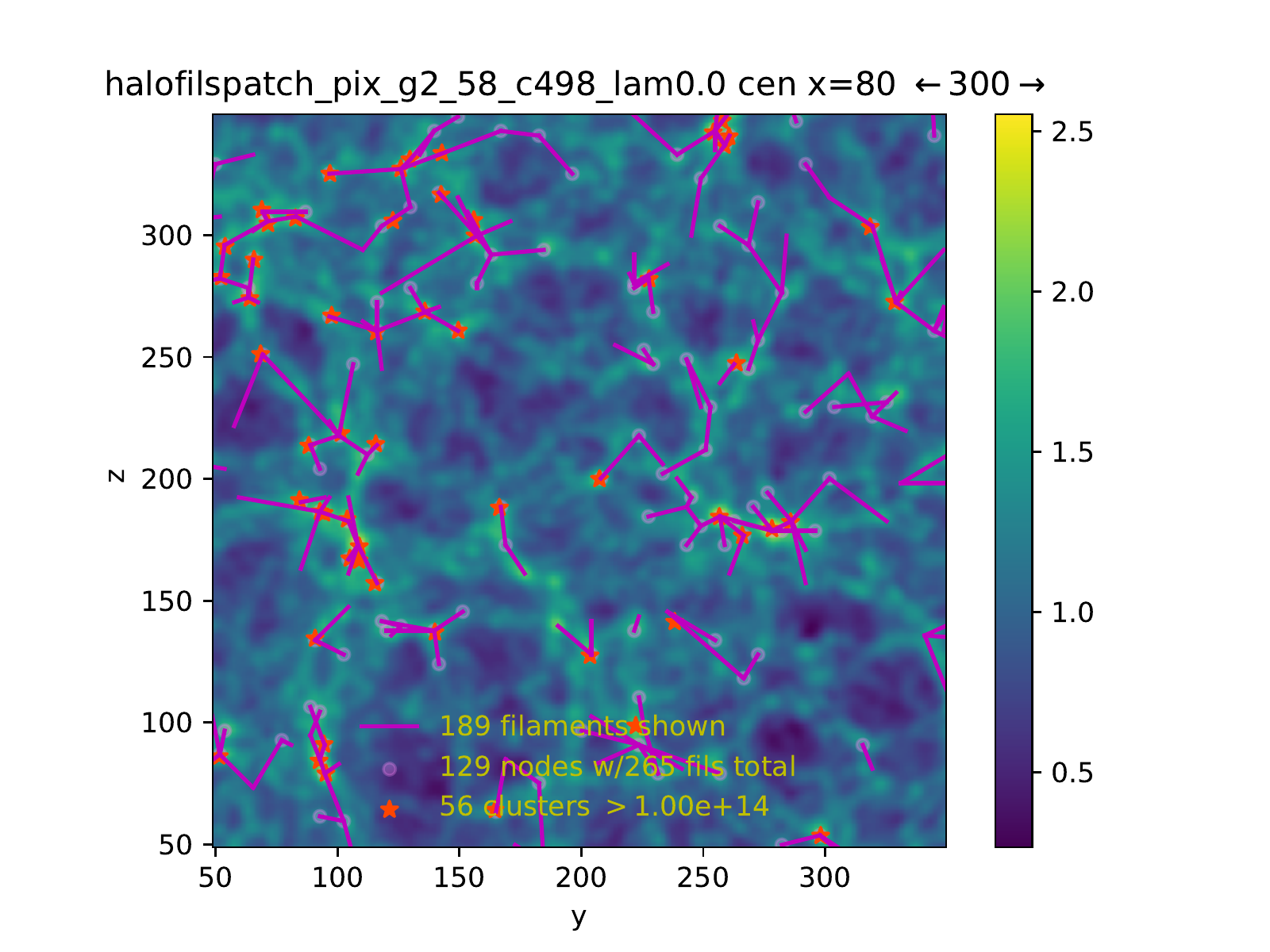}}
\resizebox{3.2in}{!}{\includegraphics{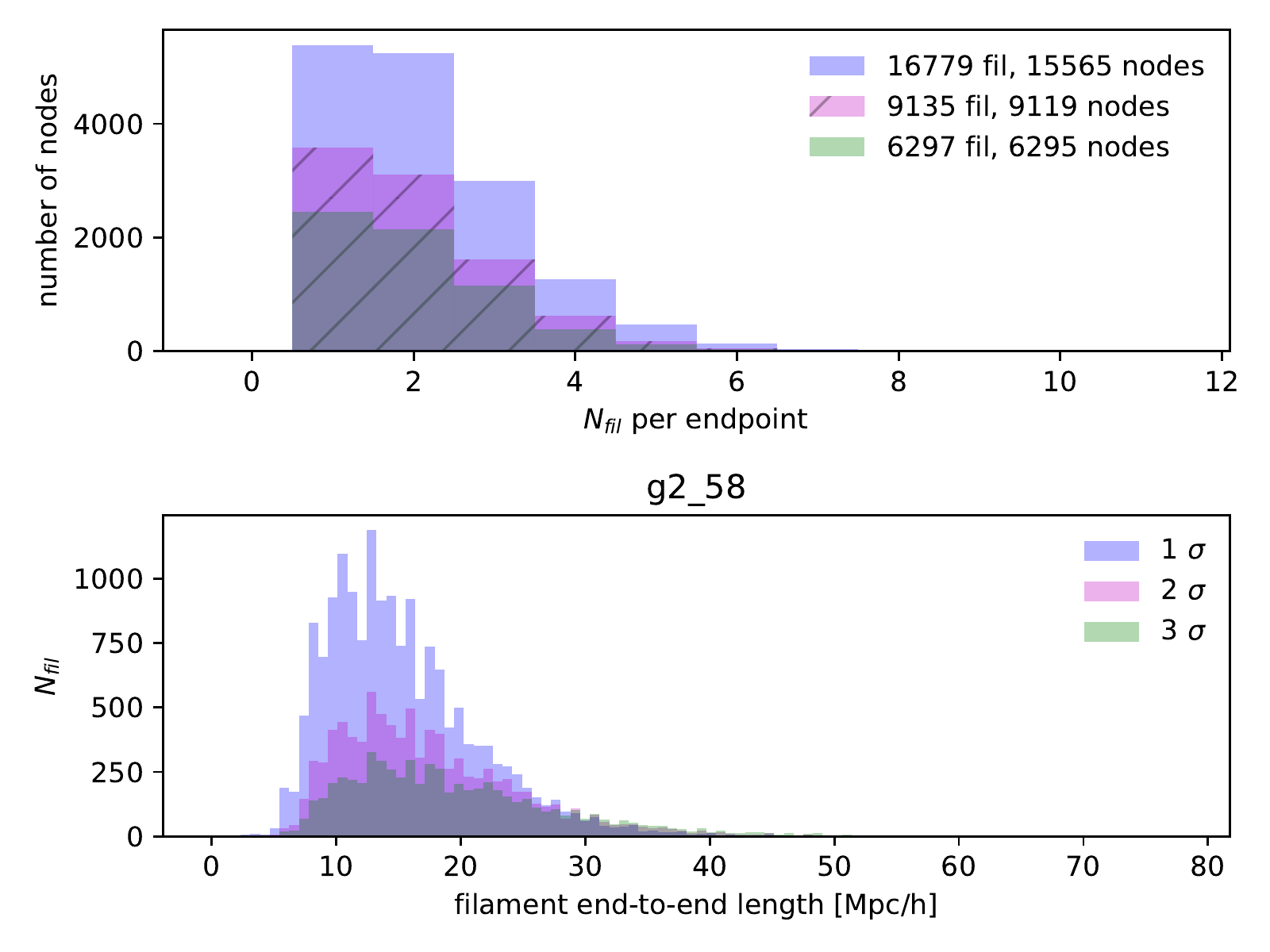}}
\end{center}
\caption{Webs with different {\sc Disperse} persistence levels.  Color
  denotes log(density).  The density slice is 30 $Mpc/h$ deep and 300 $Mpc/h$ wide, centered at position (80,200,200) $Mpc/h$, and viewed down the x axis, with smoothing 2 $Mpc/h$.
  The 3 panels show {\sc Disperse} nodes and filaments corresponding to
  persistence of 1$\sigma$ (upper left), 2$\sigma$ (upper right), 3$\sigma$ (lower left). The grey-blue dots are {\sc Disperse} node
  pixels, connected by filaments (magenta lines).  At lower right are
  statistics for the full $500 Mpc/h$ side box: the number of
  filaments per node, and distribution of filament lengths. With
  increasing persistence, the number of nodes, filaments, and
  connectivity go down, and the average filament length increases.}
\label{fig:dispall3}
\end{figure*}

\subsection{Finding the web}
Many
different web finders have been implemented and studied in past decades. The resulting different webs capture different
physical aspects and are based upon a wide range of tracers (e.g., dark matter
densities, galaxy positions, velocities, etc.) and algorithms,
e.g., eigenvalues of the shear tensor \citep[e.g.,][]{Zel70,
  Hah07a, Hah07b}, density critical points and the ridges connecting
them e.g., {\sc Disperse} \citep{Sou11,SouPicKaw11}, {\sc MMF-2}
\citep{Ara07}, {\sc NEXUS} \citep{CauWeyJon13}, shell crossing
histories and flows \citep{ShaZel89,Sha11}, phase space {\sc
  ORIGAMI} \citep{FalNeySza12,FalNey15},{\sc MWSA} \citep{RamSha15},
etc.  In \citet{Lib18}, twelve approaches were classified, reviewed, applied and
compared.  Using a common 200 $Mpc/h$ side dark
matter simulation, the webs resulting from the different approaches were compared via
volume and mass fractions for each web component,
overlaps between the same components with different finders, and more.
Webs can also be characterized by node connectivity and angular
dependence, studied, e.g., in Gaussian fields and simulations
\citep{CodPicPog18}, via histories and mergers of critical points
\citep{Cau14,Cad20}, and in terms of their relations to many different galaxy
properties \citep[there is a very long list, see, e.g.,][]{Hel21,Win21}.  Since the \citet{Lib18} comparison
paper, methods for web detection \citep{Fan19,Per20,Wan20} and web
finder comparison methods \citep[e.g., between filaments,][]{Ros20} have
continued to be developed.  Each web finding method involves choices
of parameters, scales and (sometimes implicitly) other assumptions.

The main web finder used here is {\sc Disperse}
\citep{Sou11,SouPicKaw11}\footnote{http://www2.iap.fr/users/sousbie/web/html/indexd41d.html
}, a publicly available code which does a multiscale identification of
all 4 web components starting with individual objects (such as
galaxies) or a density field.  Here, the (Gaussian) smoothed density
field is taken to be the starting point and {\sc Disperse} is used to identify
nodes and filaments from pixel maps of simulation densities relative
to the mean.\footnote{The added level of complication for {\sc Disperse} on
  subhalo counts became unmanageable
  in practice for the available computing resources and current data
  set.}  Using Morse theory, {\sc Disperse} classifies regions using
critical points (nulls of the gradient of the field) and integral
lines (tangents to the field at every point, which converge at
critical points).  In addition to the underlying smoothing and pixel size, the {\sc Disperse} web also depends on
``persistence'', a parameter describing roughly how much a field has
to decrease between two peaks in order for them to be considered as
distinct peaks.  As the persistence rises, {\sc Disperse} nodes are removed.  Persistence levels in the examples below are taken to
be 1$\sigma$, 2$\sigma$ and 3$\sigma$ of the density field; although
$1\sigma$ is a low persistence choice, it helps in examining trends
with persistence.\footnote{ To run, {\sc Disperse} uses a file of pixel
  density values, e.g., ``pixfile.dat''. One chooses persistence
  ``pers'', which is a number, and runs {\sc mse pixfile.dat -cut pers
    -upSkl}.  On that output.up.NDskl one runs {\sc skelconv
    output.up.NDskl -Outname pixfile -to NDskl\_ascii } to get
  pixfile.a.NDskl, with node and saddle (filament center) critical
  points and information about each.} 

With the {\sc Disperse} webs, the nodes are points (dimension
zero)\footnote{The allowed {\sc Disperse} node positions on a pixel
  are at (0, 1/4, 1/3, 1/2, 2/3, 3/4) in pixel units.  The pixel size
  500/256 $Mpc/h$ is used to rescale, and then {\sc Disperse} provided
  positions are shifted by half a pixel.}  and the filaments are curves (dimension 1) associated with saddle critical points (dimension 0), none of which have associated volume.  One way
to associate clusters to a web node is to have the cluster lying
``in'' the node. Two obvious scales for assigning volume to {\sc
  Disperse} nodes, within which a cluster could lie, are the pixel
size and smoothing scale.  A third way to associate volume to a {\sc
  Disperse} node, which uses more information from the density field,
is based upon the Hessian based web finder, or ``T-web''; some recent
descriptions are in \citet{Hah07a,Hah07b}.  (Other ways of possibly
defining the ``width'' of a {\sc Disperse} node or filament, beyond
what will be considered here, include stacking {\sc Disperse} objects
\citep{Kra19} or looking at vorticity behavior \citep{Lai15}.)  To
classify pixels in terms of cosmic web components, the Hessian based
web finder uses eigenvalues $\lambda_1\geq \lambda_2\geq \lambda_3$ of
the shear tensor $T_{ij} =\partial_i \partial_j \phi$, where $\phi$ is
the gravitational potential \citep{Zel70}.  The eigenvalues are
calculated for each pixel from the smoothed density field.  Every
pixel is then assigned to a web component based on how many
eigenvalues are $>$ or $<$ 0, nodes correspond to ($+++$), filaments to
($++-$), walls to ($+--$) and voids to ($---$). One can generalize to
classifying eigenvalues being above or below \citep{For09} some value
$\lambda_{th}$, which then becomes an additional web parameter.  To
construct objects such as a node or filament from these individually
labelled pixels, more assumptions are needed. In the following, every
set of contiguous node pixels will be considered a separate Hessian
``node'' or ``patch.'' These patches will be used in one of the cluster-{\sc
  Disperse} node matching methods below (which assigns Hessian patch regions to
{\sc Disperse} nodes lying within them). Even more assumptions
would be required to identify and construct Hessian based filament
objects; this will not be pursued here.\footnote{An example of a
  method to construct filament objects from a Hessian description is found in \citet{Pfe22}.}

Again, both the {\sc Disperse} and Hessian web classifications depend upon
the smoothing length and pixel scale, parameters not intrinsic to the
mass distribution itself, {\sc Disperse} also depends upon persistence and the Hessian classification also depends upon $\lambda_{th}$.

A 30 $Mpc/h$ deep simulation slice is shown in
Fig.~\ref{fig:dispall3}, with {\sc Disperse} nodes and
filaments, for persistence 1$\sigma$, 2$\sigma$ and 3$\sigma$ (the smoothing is 2 $Mpc/h$).  For these three {\sc Disperse} webs, the distribution of {\sc Disperse} node and filament counts,
node connectivities, and filament lengths are intercompared in the box at
lower right.  The $3 \sigma$ persistence {\sc Disperse} web with smoothing 2$Mpc/h$
will be used below when examples for a single parameter choice are shown.  The 2 $Mpc/h$ smoothing is motivated by its frequent use
\citep[e.g.,][]{Hah07a, Hah07b}.  

\begin{figure}  
\begin{center}
\resizebox{3.5in}{!}{\includegraphics{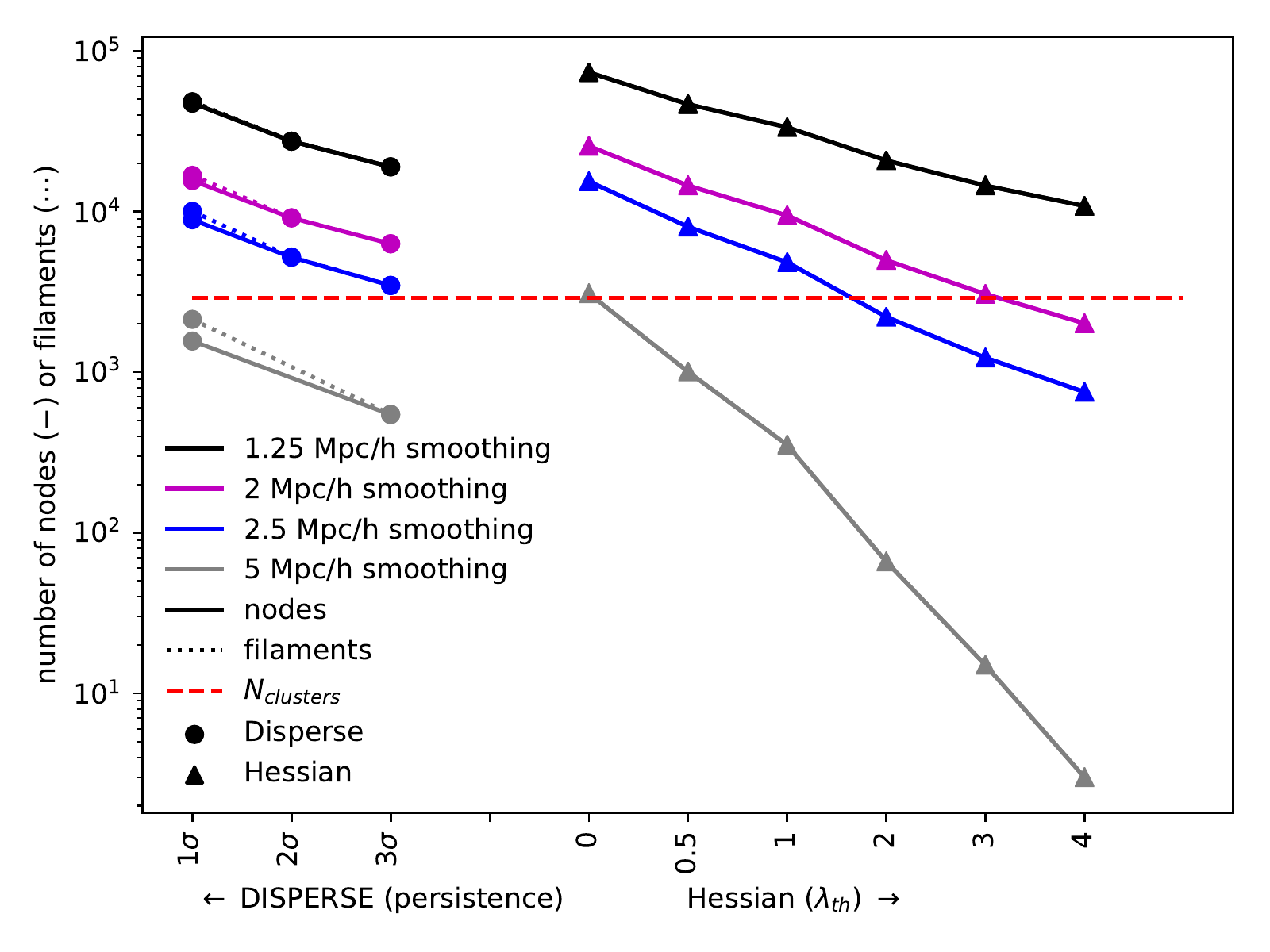}}
\end{center}
\caption{Node counts in different web descriptions, relative to the number of
  clusters, the 2898 massive halos with $M\geq 10^{14} M_\odot$ (red dashed
  line). For each smoothing (color), the left 3 points are the number
  of {\sc Disperse} nodes (circles), ordered by increasing persistence.  The
  right 5 points are the number of Hessian nodes (triangles), labeled
  by $\lambda_{th}$ (in units of $\partial_i \partial_j\nabla^{-2}
  \rho/\bar{\rho}$). Dotted lines show the number of {\sc Disperse}
  filaments, close to the number of nodes except for the $1 \sigma$ persistence, $5 Mpc/h$ smoothing.}
\label{fig:numnodes}
\end{figure}
\section{Associating clusters and nodes}
\label{sec:associate}
\subsection{Nodes: counts and Hessian node patch properties}
Before associating clusters and nodes, the first question is how
many of each are present?  Node counts are summarized for several Hessian and
{\sc Disperse} choices of parameters in Figure \ref{fig:numnodes}.
The numbers of nodes are shown at left by circles for {\sc Disperse}
persistence of $(1 \sigma, 2 \sigma, 3 \sigma)$ and at right by
triangles for Hessian thresholds $\lambda_{th}$ = (0, 0.5, 1, 2, 3, 4).
Color denotes the different (1.25, 2.0, 2.5, 5.0) $Mpc/h$ smoothings.  The red
dashed line is the number of clusters.  

For all the smoothings considered, not just the 2 $Mpc/h$ example in Fig.~\ref{fig:dispall3}, as {\sc Disperse}
persistence goes up, the number of nodes and filaments go down, and longer filaments become a higher fraction of all
filaments (as the nodes become rarer and further apart).
Often the number of filaments does not exceed the number of nodes by
much, see Fig. ~\ref{fig:numnodes}.\footnote{Within 0.4\% except for $1 \sigma $ persistence,
where $N_{node}/N_{fil}$ drops from 98\% to 73 \% as smoothing increases from
1.25 $Mpc/h$ to 5 $Mpc/h$.} 

Similarly, the number of Hessian node patches decreases
as the threshold $\lambda_{th}$ eigenvalues must cross, in order
to qualify as a node, increases.  As mentioned earlier, {\sc
  Disperse} nodes are single points in a single pixel,
parameterized by position and density. The Hessian node patches have,
in addition, a size, that is, the number of continguous pixels classified as part of a given
node. Larger Hessian patches are less frequent as the threshold
$\lambda_{th}$ increases and more frequent as smoothing increases.\footnote{For
$\lambda_{th}=0$ and smoothing 1.25 $Mpc/h$, 40\% of the Hessian nodes
have only one pixel, while for the more restrictive $\lambda_{th} =
4$, 71\% of the Hessian nodes only have one pixel.  As smoothing
increases, fewer node patches have only one pixel, e.g., only 3\% do
so for smoothing 5 $Mpc/h$ and $\lambda_{th} = 0$.}
\begin{figure}
\begin{center}
\resizebox{3.5in}{!}{\includegraphics{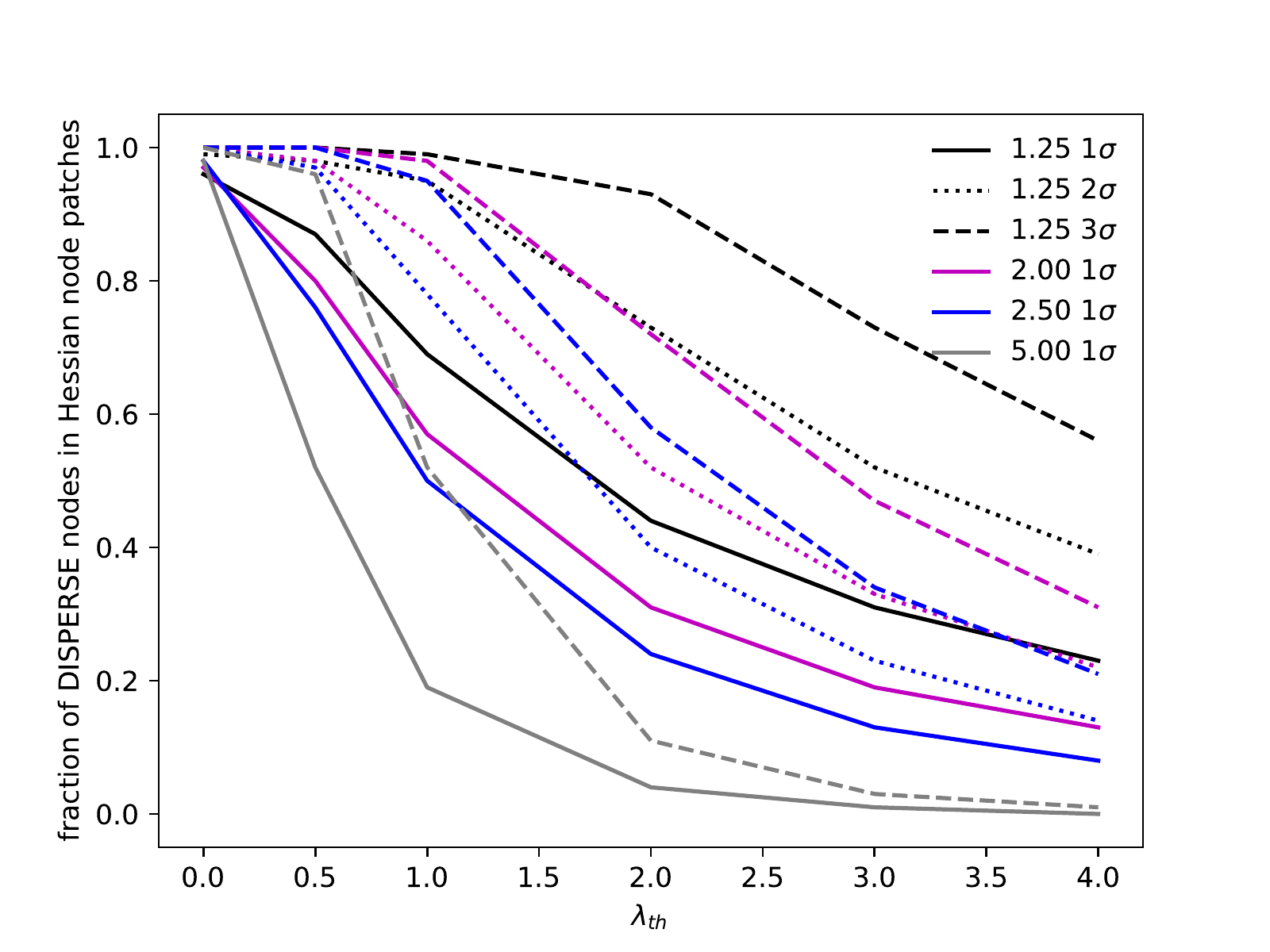}}
\end{center}
\caption{{\sc Disperse} nodes and Hessian node patches: the fraction of {\sc Disperse} nodes which have a
  corresponding Hessian node (via lying in a shared pixel), for {\sc
    Disperse} persistence: $1\sigma$ (solid), $2\sigma$ (dotted),
  $3\sigma$ (dashed).  The color denotes Gaussian smoothing 
  (1.25 $Mpc/h$, 2  $Mpc/h$, 2.5 $Mpc/h$, 5 $Mpc/h$) as indicated.
  For fixed $\lambda_{th}$, higher persistence and lower smoothing
  webs give a higher fraction of
  {\sc Disperse} nodes matched to Hessian nodes..}
\label{fig:dispinhess} 
\end{figure}

\begin{figure}
\begin{center}
\resizebox{3.5in}{!}{\includegraphics{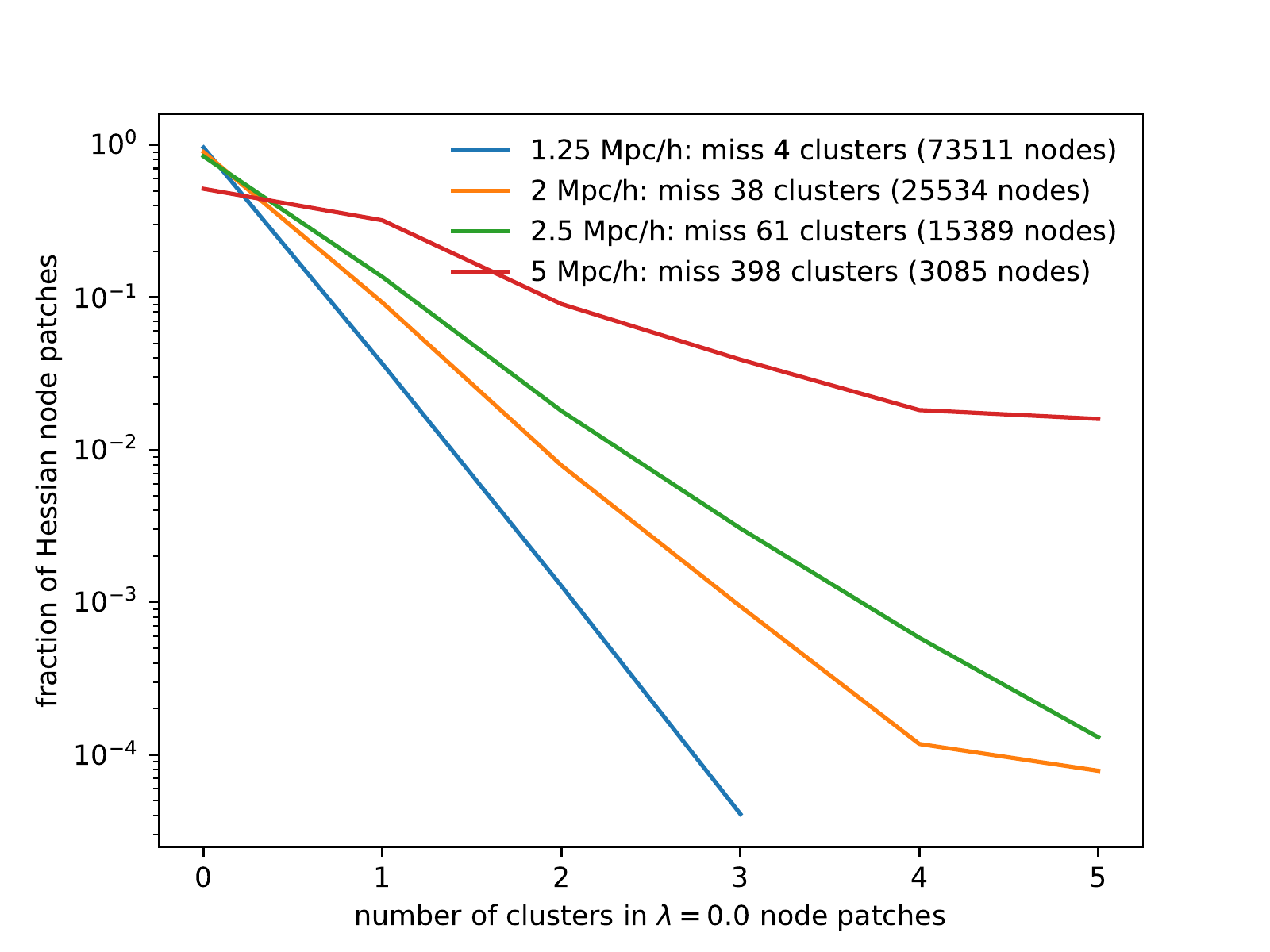}}
\end{center}
\caption{Clusters and Hessian node patches:  the fraction (note the
  log scale) of $\lambda_{th}=0$ Hessian node patches 
which contain a given number of clusters, as a function of
Hessian web smoothing, or ``node occupation distribution.'' The
Hessian patch nodes significantly outnumber the clusters except for
the 5 $Mpc/h$ smoothing. The numbers of
clusters (out of 2898) not lying in any Hessian node patch, for each smoothing, are shown in the legend.}
\label{fig:clusinhess} 
\end{figure}

Looking ahead to matching clusters and {\sc Disperse} nodes lying in
the same Hessian node patch, in Fig.~\ref{fig:dispinhess} the fraction of {\sc Disperse}
nodes which lie within (pixels which are part of) Hessian node patches is shown as a function of threshold
$\lambda_{th}$ and {\sc Disperse} persistence.  As the {\sc Disperse}
persistence is raised, a higher fraction of {\sc Disperse} nodes lie
in Hessian node pixels for a given $\lambda_{th}$, that is, the
definitions tend to overlap more often.  Given the steep drop in the
number of matched {\sc Disperse} nodes to Hessian patches as
$\lambda_{th}$ is raised, in order to have more {\sc Disperse} nodes
available to match to clusters, only $\lambda_{th}=0$ Hessian patches
will be used hereon.  As smoothing increases, a smaller fraction of
{\sc Disperse} peaks lie in Hessian patches for a given $\lambda_{th}$.

Turning to clusters, most Hessian nodes have no clusters lying inside them at all.
However, because the Hessian patches can be large, in some cases 
with average radii from the maximum density point of 12-16 $Mpc/h$, a
fraction of Hessian node patches contain several clusters.  This allows one to
construct a ``node
occupation distribution'' (by clusters), shown for the four different Hessian web smoothings
(again, $\lambda_{th}=0$) in Fig.~\ref{fig:clusinhess}.
That is, this is the fraction of Hessian node patches containing a
given number of clusters (ranging from 0 to 5, except for 5 $Mpc/h$ smoothing, with a maximum of 12). More clusters share Hessian patches (i.e.,
are in the same Hessian patch, together) and more clusters are missed, as the smoothing
increases (legend, Fig.~\ref{fig:clusinhess}).

\subsection{Associating clusters to nodes}
\label{sec:clus2node}
\begin{figure}
\begin{center}
\resizebox{3.5in}{!}{\includegraphics{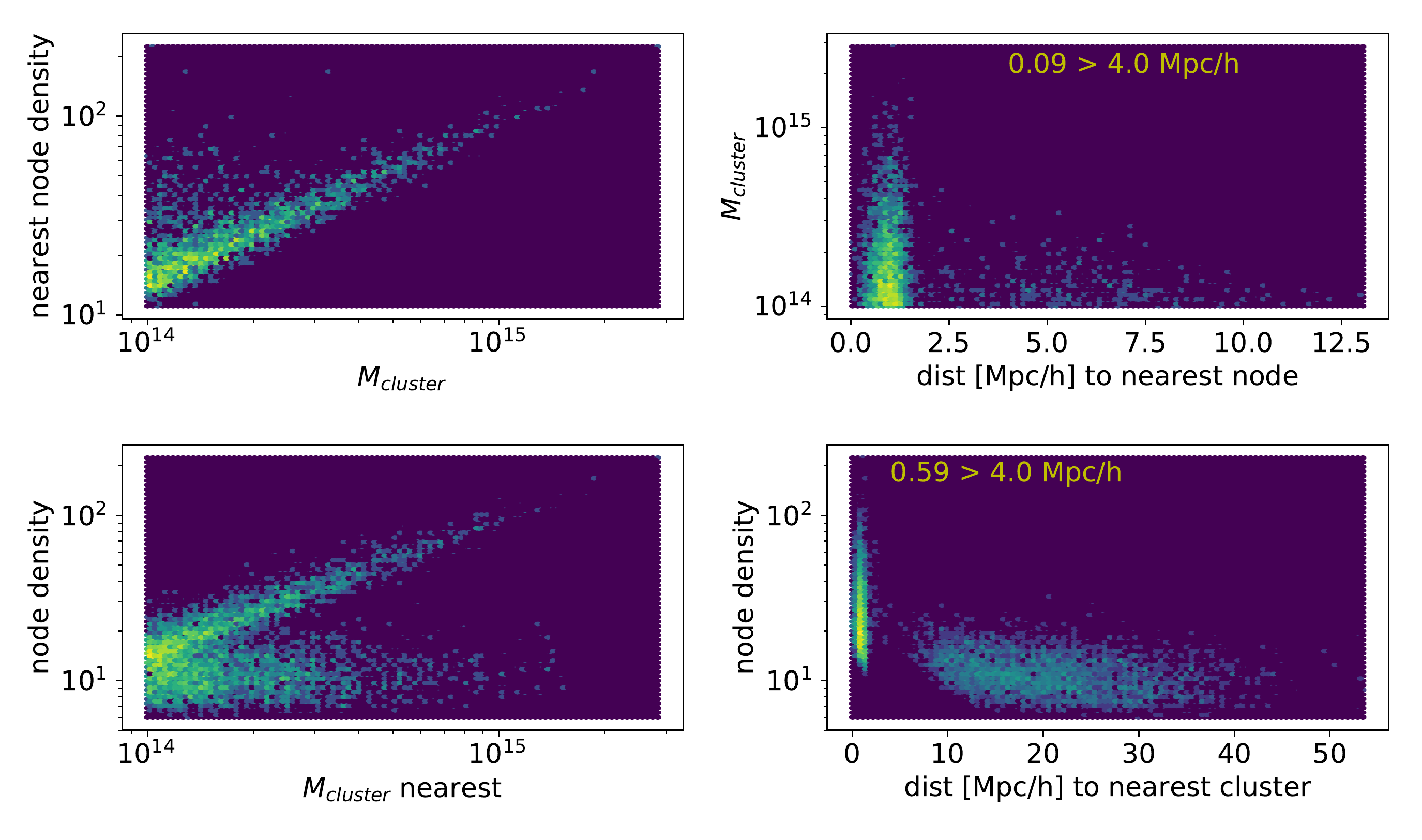}}
\resizebox{1.63in}{!}{\includegraphics{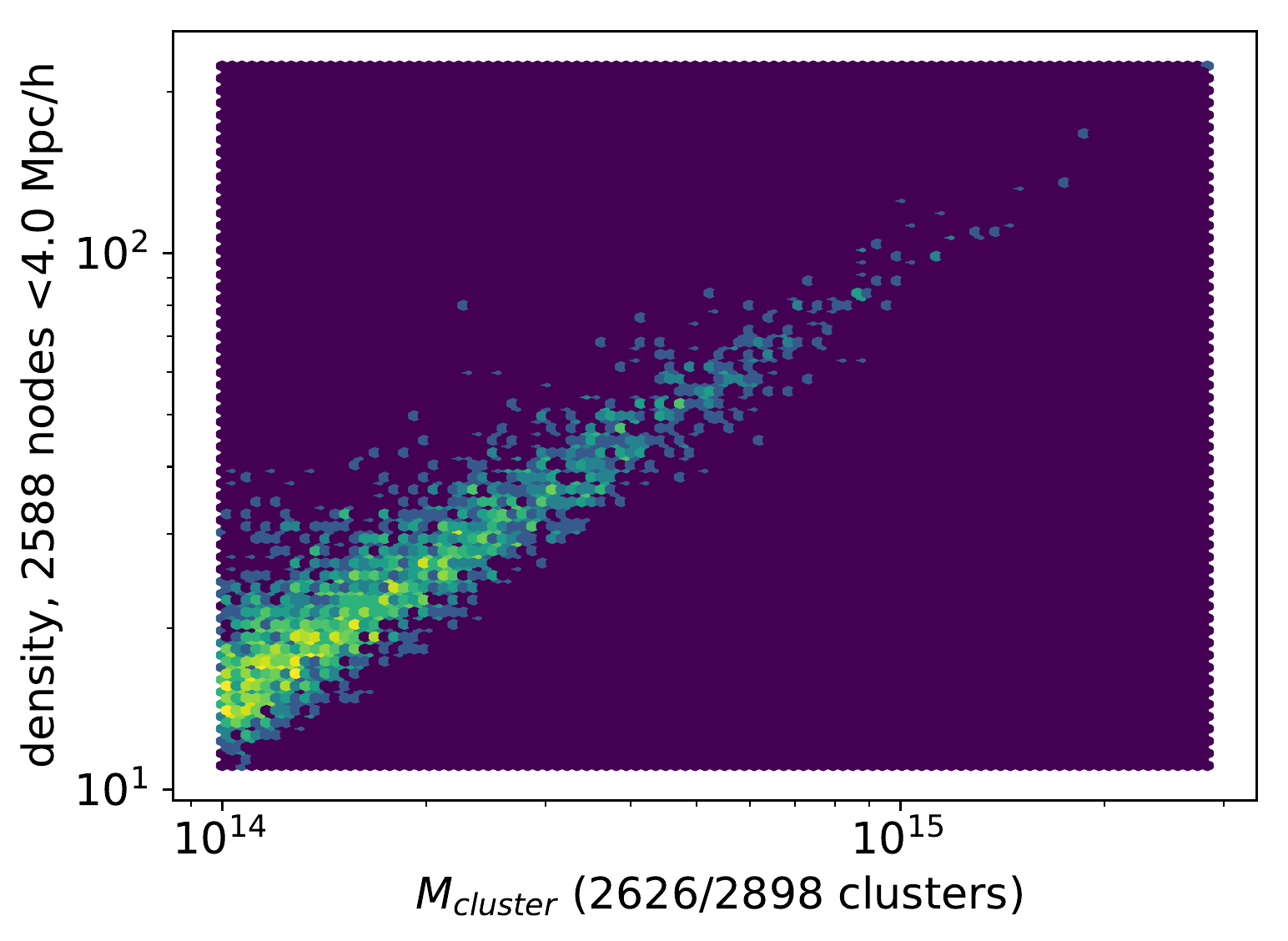}}
\resizebox{1.63in}{!}{\includegraphics{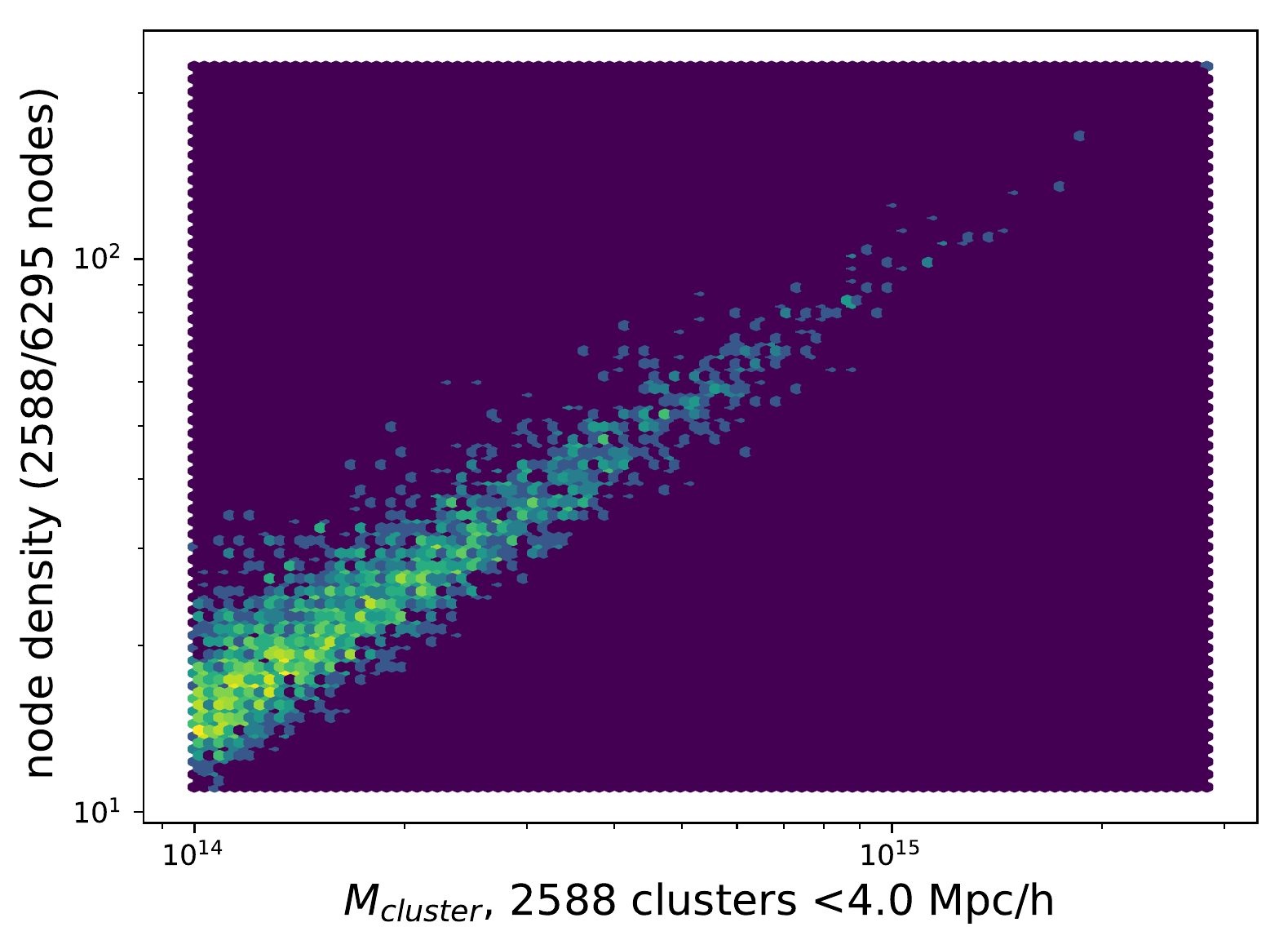}}
\end{center}
\caption{Relation between clusters and their nearest {\sc Disperse}
  node, for $3\sigma$ persistence and smoothing 2 $Mpc/h$.
At top, the nearest node for each of the 2898 clusters is shown.  The
  low mass clusters are paired with nodes having a range of densities,
  and sometimes that association is weak, as the nearest node can be
  up to 15 $Mpc/h$ away.  Within twice the smoothing distance (4
  $Mpc/h$), 9\% of clusters do not have a {\sc Disperse} node. At
  middle, the nearest cluster for each of the 6295 {\sc Disperse}
  nodes is shown (there are many more nodes than clusters), which are sometimes $>$ 40 $Mpc/h$ away.  59\% of the 6295
  {\sc Disperse} nodes do not have a cluster within 4 $Mpc/h$. The
  lowest density nodes have a wide range of paired cluster masses.  More
  than one cluster can have the same nearest {\sc Disperse} node and
  vice versa.  The color scheme for densities in each plot pixel is
  logarithmic.
  At bottom, by restricting to {\sc Disperse} node-cluster
  pairs within twice the smoothing length, a cluster mass-{\sc
    Disperse} node density relation is seen. At left, fixing the 2898
  clusters and finding the closest {\sc Disperse} node, at right, vice
  versa.}
\label{fig:nodeclussep}
\end{figure}

A correlation between clusters and {\sc Disperse} nodes is expected, but
the relation is not necessarily expected to be one to one.  In particular,
the {\sc Disperse} web finds critical points which are peaks (nodes, with a height difference requirement depending on persistence) and saddle points in ridges (filaments), and depends upon smoothing and pixel scales as well.
Clusters have integrated density above some absolute scale (mass
scale), and do not depend upon the smoothing and pixel scales.

Several ways to associate clusters to the {\sc Disperse} nodes were
considered and are compared below:
\begin{itemize}
\item ``fixed'' and ``nearest'': if, for a given {\sc Disperse} node, the cluster center is within some
  fixed distance of (``fixed''), or the cluster center is
  the nearest cluster center (``nearest'') within that fixed distance;
  the fixed distance is taken
  to be twice the smoothing 
\item ``pix'': if the cluster center and the {\sc Disperse} node lie in the
  same pixel 
\item ``patch'': if the cluster center pixel is in the same Hessian node patch as the {\sc Disperse} node pixel 

\end{itemize}

Other matching methods using Hessian patches were considered, such as
clusters being within a set distance
from a Hessian node patch center or peak, or within a set distance from any pixel in a Hessian node
patch, but all of these introduce more complexity, assumptions and
difficulty in interpretation.

For matching clusters to {\sc Disperse} nodes based upon distance
(``fixed'' and ``nearest''), the choice of distance scale was guided by looking at the distance to the nearest {\sc Disperse}
node for each cluster (e.g., Fig.~\ref{fig:nodeclussep}, upper
right) and the distance to the nearest cluster for each {\sc Disperse}
node
(Fig.~\ref{fig:nodeclussep}, middle right), shown for the reference {\sc Disperse} web with 2
$Mpc/h$ smoothing and $3\sigma$ persistence. For all smoothings,
there seems to be a clear subset of clusters with a {\sc Disperse} node
close by, roughly captured by using twice the smoothing scale as a
distance cutoff.\footnote{The pixel scale of $\sim$ 2 $Mpc/h$ seems relevant, as
{\sc Disperse} nodes are only defined on a grid with that resolution, i.e.,
{\sc Disperse} node location changes on smaller scales do not reflect
the underlying matter distribution on smaller scales.  It was not
clear how to take this into account.}  
Restricting to only {\sc
  Disperse} node-cluster pairs within twice the smoothing length, a
{\sc Disperse} node density-cluster mass relation is seen all the way down to the lowest {\sc Disperse} node
densities and cluster masses(Fig.~\ref{fig:nodeclussep}, bottom).

More than one cluster can have the same nearest {\sc Disperse} node and vice versa. The number of times the closest {\sc  Disperse} node to a cluster within the (twice smoothing) distance
cut is already matched is $<$25 for the 1.25 $Mpc/h$ smoothing,
dropping to 5 additional clusters for 2 $Mpc/h$ smoothing with 1$\sigma$
persistence and then becoming zero for larger smoothings or
persistence.  In contrast, the number of {\sc Disperse} nodes whose
nearest cluster is already matched rises with smoothing
($\sim 1, \sim 40, \sim 110, >300$) for smoothings (1.25 $Mpc/h$, 2
$Mpc/h$, 2.5 $Mpc/h$, 5 $Mpc/h$).  
Both ``fixed'' and ``patch'' also do not necessarily
associate a unique cluster to a given {\sc Disperse} node (or a unique {\sc Disperse} node to a given cluster, for patch), as they map
all clusters in a given region to the same {\sc Disperse} node. 

The pixel matching method is unambiguous but can
fail to match a cluster to a {\sc Disperse} node if the cluster
center is offset from the pixel containing the latter, even if the
cluster itself extends into the {\sc Disperse} node pixel.  So
although clearly defined, there are reasons to think it might be too restrictive.

\begin{figure}
\begin{center}
\resizebox{3.5in}{!}{\includegraphics{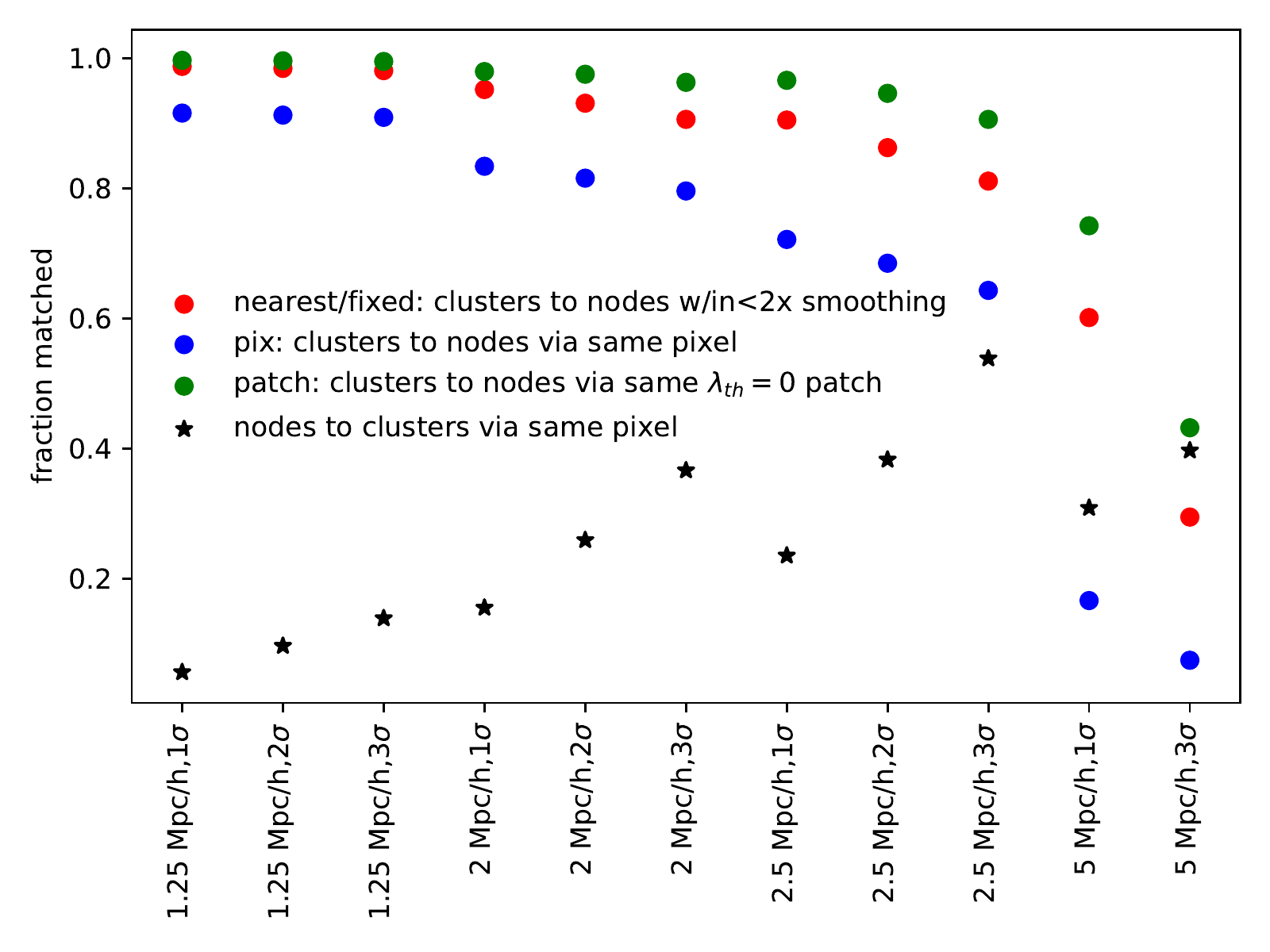}}
\end{center}
\caption{Fractions of clusters matched to {\sc Disperse} nodes via the
  ``nearest/fixed'', ``pix'', ``patch'' methods, as a function of
  smoothings and persistence ($1\sigma$, $2\sigma$, $3\sigma$) and methods; clusters
  lying in the same $\lambda_{th}=0$ Hessian patch (green dots,
  ``patch''), lying within a distance of twice the smoothing
  length (red dots, ``nearest'' or ``fixed''), and lying within the
  same pixel (blue dots,``pix'') as a {\sc Disperse} node. The
  $\lambda_{th}=0$ Hessian patch and the {\sc Disperse} $1 \sigma$
  persistence 1.25 $Mpc/h$ smoothed nodes gives the most matched
  clusters (2889/2898).  The ``patch'' method for higher
  $\lambda_{th}$ values has fewer clusters matching {\sc Disperse}
  nodes relative to all 3 methods shown.  Stars refer to the fractions of
  {\sc Disperse} nodes with clusters in the same pixel, matching via the ``pix'' method, a similar fraction (from 94\% to 29\%) of {\sc
    Disperse} nodes do not have a cluster within 2 $Mpc/h$.}
\label{fig:halonodepix}
\end{figure}

Although for almost all {\sc Disperse} webs there are many more {\sc
  Disperse} nodes than clusters, not all clusters are matched to {\sc
  Disperse} nodes.  The fraction of clusters which are
matched to {\sc Disperse} nodes by
each of the 4 methods above,\footnote{Note that ``nearest/fixed'' are two methods, in one case only one {\sc Disperse} node is matched, in the other, all {\sc Disperse} nodes within twice the smoothing are matched.  For counting how many clusters have at least one match, these give the same result, and so are shown as the same point.  However, the two cases correspond to different configurations more generally, so are counted separately, i.e. 36 webs are considered, in Fig.~\ref{fig:missclus} below.} for the different {\sc Disperse} webs, is
shown in Fig.~\ref{fig:halonodepix}.\footnote{One way of evaluating this matching is to compare how well finding {\sc Disperse} nodes manages to find clusters, to how well different halo finders find the same object (cluster).  In Fig.~17 of
  \citet{Kne11}, a halo finder comparison paper, the number of halos above $10^{14} M_\odot$ can differ by around 10\%  between some finders, comparable to or larger than the number of clusters missed by the ``nearest/fixed'' and ``patch'' methods in Fig.~\ref{fig:halonodepix}, except for the largest 5 $Mpc/h$ smoothing and the 2.5 $Mpc/h$ smoothing with 2$\sigma$ or 3$\sigma$ persistence.  I thank C. Miller for suggesting this comparison and the reference to help make the comparison.}    In contrast to the large number
of matched clusters, most of the {\sc
  Disperse} nodes are not matched to clusters (for example, stars in
Fig.~\ref{fig:halonodepix}, for ''pix'' matching), although going to low enough halo or even subhalo mass would perhaps give at
least one (sub)halo in the same pixel as every node.  Increasing the fraction of matched
{\sc Disperse} nodes by raising persistence or smoothing causes the
number of matched clusters to decrease.

\begin{figure}
\begin{center}
\resizebox{3.5in}{!}{\includegraphics{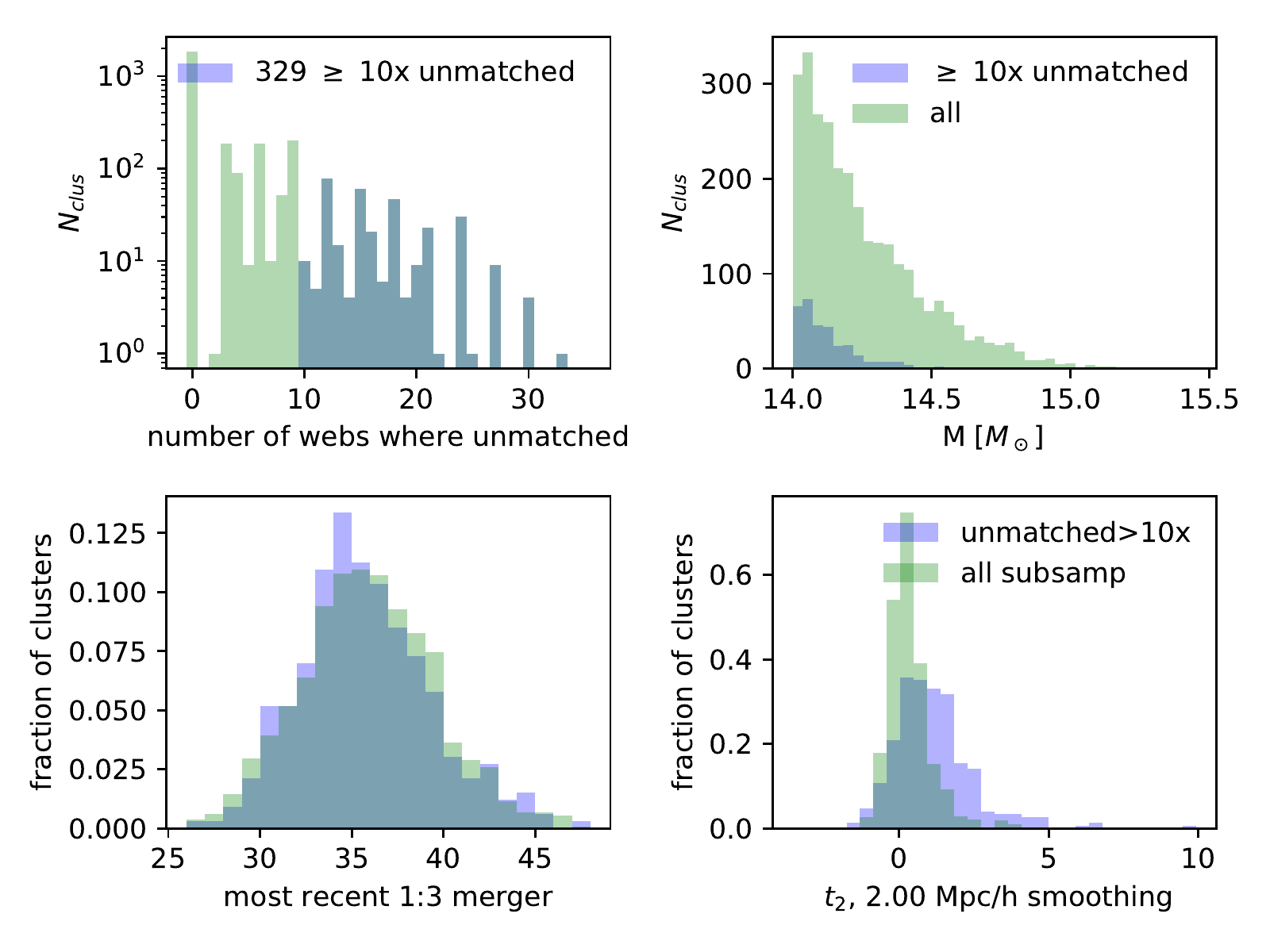}}
\end{center}
\caption{Properties of clusters, shown in blue, which are unmatched in 10 or more of the 36
  combinations of {\sc Disperse} webs and cluster matching methods
  (smoothing $<$ 5 $Mpc/h$).  Upper left is the number times each
  cluster is unmatched for the 36 combinations (``nearest'' and ``fixed'' are degenerate for this quantity).
  Upper right shows the full
  cluster mass distribution; the unmatched cluster
  mass distribution tends to be lower mass.  The lower panels
  compare a mass matched sample of all clusters (lighter green) with the same 329 clusters unmatched for $\geq$ 10 of the 36 combinations (in blue).  At lower left,  the
  most recent 1:3 merger time step is perhaps slightly less recent for
  unmatched clusters (more recent is higher number).  At  lower
  right,  $t_2 = \lambda_2 - \delta/3$ (calculated for smoothing 2
  $Mpc/h$) seems higher for the unmatched clusters.
  Similar trends were seen for the (smaller number of) clusters which
  remained frequently unmatched when considering only matching via  ``nearest'', ``fixed'' and ``patch'' (i.e., dropping ``pix'').}
\label{fig:missclus}
\end{figure}

\begin{figure}
\begin{center}
\resizebox{3.5in}{!}{\includegraphics{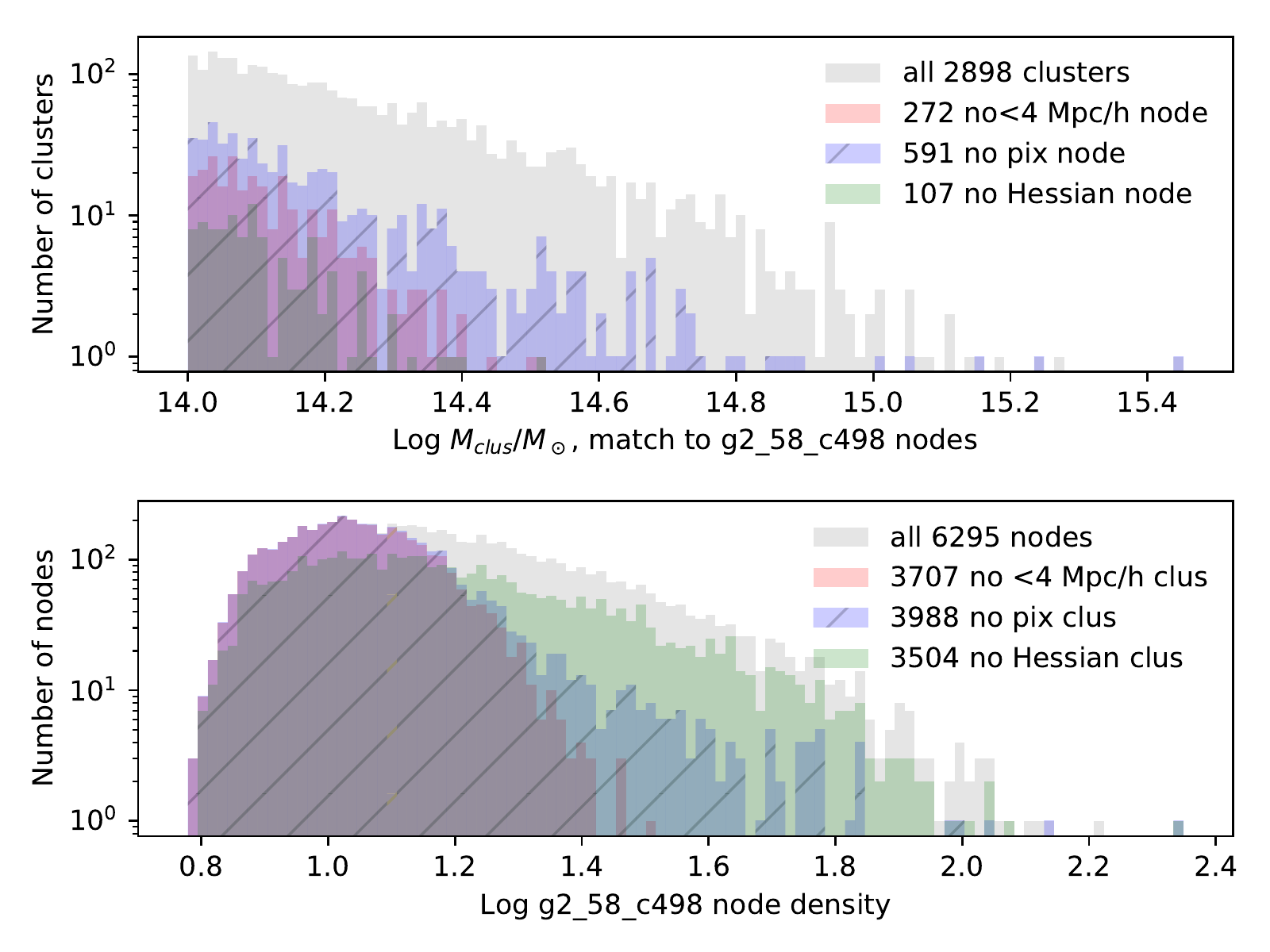}}
\end{center}
\caption{Cluster mass function (top) and {\sc Disperse} node density function (bottom), grey, for the reference 3$\sigma$ persistence $2 Mpc/h$
  smoothed {\sc Disperse} web.  In color are properties of clusters unmatched to {\sc Disperse} nodes (top) and vice versa (bottom), via three matching methods. The three methods are lying within twice the smoothing length (4 $Mpc/h$, ``nearest''),
   lying in the same pixel (``pix''), or lying in
  the same Hessian node patch (``patch''). Unmatched clusters (top) 
  missed by the ``patch'' and ``nearest'' methods
  seem to lie in the low mass range, while the many more clusters missed
  by the ``pix'' method ranged all the way to the highest mass.  In comparison, almost all the low density {\sc Disperse}
  nodes are missed by the ``nearest'' and ``pix'' method, but {\sc Disperse} nodes missed by the ``pix'' method also go up to higher density.  The ``patch'' method has fewer {\sc Disperse} nodes unmatched to clusters, but many more of the ``patch'' unmatched {\sc Disperse} nodes are at high density.
}
\label{fig:missclusnodes}
\end{figure}

Setting aside the two 5 $Mpc/h$ smoothed webs
for the rest of this section, as their {\sc Disperse} nodes miss
significantly more clusters, there are 36 combinations of {\sc
  Disperse} webs and matching methods.  For these, a histogram of the number of times each cluster is unmatched (from 0 to 36 out of the 36 combinations) is shown in the upper left hand corner of
Fig.~\ref{fig:missclus}.  Over half of the clusters (1832/2898) were
matched to {\sc Disperse} nodes for every web and method.  Dropping
the ``pix'' method, which contributes 9 of the 36 variations, and is the most restrictive way to match
clusters and {\sc Disperse nodes}, 2248/2898 clusters, about 3/4, were always
matched to {\sc Disperse} nodes, by all remaining 3 methods.

One cluster was missed by 
every method, by dint of being slightly beyond the distance cutoff (of twice the
smoothing scale), while 79 (329) were not matched to {\sc Disperse}
nodes for 20 (10) or more web/method
combinations (out of 36).   Some characteristics of the 329 clusters which were unmatched for 10
or more of the 36 web/matching combinations are shown in 
Fig.~\ref{fig:missclus}.  At upper right is their mass distribution
(they tend to be lower mass). At lower left is the time of their most
recent major (1:3) merger, which seems slightly earlier than for a
random cluster
subsample with the same mass distribution.  For
different underlying {\sc Disperse} webs, the merger history
difference is most pronounced for the smallest smoothing and increases
as the maximum allowed distance for matching between clusters and
{\sc Disperse} nodes increases from twice the smoothing scale.

At lower right is the distribution, for the frequently
unmatched clusters versus a random cluster subsample with the same mass distribution, of the middle component of ``velocity shear'' \citep{LudPor11},
$t_2 = \lambda_2 - \delta/3$.   The $t_2$ distribution of the frequently unmatched clusters
seems to be higher relative to the subsample of all clusters.  (When
considering clusters unmatched for a specific {\sc Disperse} web matching,
 $t_2$ is calculated using the same smoothing
scale as the {\sc Disperse} nodes.  For this figure, 2 $Mpc/h$ is used.)
The ``velocity shears'' $t_i$,  refer to
whether local collapse is favored (if $>0$) or impeded (if $<0$) along
a given axis, and these unmatched clusters seem to be in regions that
have collapse favored along two axes more frequently ($t_2>0$) relative to all clusters.  More frequent $t_2>0$ was also
found for the initial conditions of ``peakless halos'',
massive halos lacking a corresponding high mass peak in the initial conditions
\citep{LudPor11}.  
The difference in $t_2$ between matched and
unmatched clusters is largest for webs with smaller smoothing and for matching
methods which produce fewer unmatched clusters.  Samples of
clusters which are further away from the nearest
{\sc Disperse} nodes tend to have the highest
difference in $t_2$ from the matched clusters.

For {\sc Disperse} nodes, those which are unmatched  tend (slightly)
to be lower density (an analogue of mass)
for some smoothings and persistences, but it is harder to look at
trends across different webs as the {\sc Disperse} nodes themselves also vary between webs. 

Looking at a single example, the reference {\sc Disperse} web (3$\sigma$ persistence and 2 $Mpc/h$ smoothing),  Fig. ~\ref{fig:missclusnodes} shows the mass distributions for unmatched clusters (top)  and density distributions for unmatched {\sc Disperse} nodes (bottom), as the cluster-{\sc Disperse} node matching methods are varied.  The mass
distribution of unmatched clusters seems similar for the three matching methods, with the method missing the most clusters extending to the highest mass clusters. In contrast, the density of unmatched {\sc
  Disperse} nodes seems significantly different for the ``patch''
method compared to the other two, in particular, the ``patch'' (i.e.,
Hessian) method does match some of the low density {\sc Disperse}
nodes to clusters, while the other two methods seem to result in almost all of the low density {\sc Disperse} nodes being missed.

In summary, {\sc Disperse} nodes can be matched to clusters in several
ways.  For the 9 {\sc Disperse} webs, the 4 matching methods, and smoothing $< 5 Mpc/h$, over
half of the clusters always have a matched {\sc Disperse} node.  Over 3/4 of the clusters have a {\sc Disperse}
node match if the ``pix'' matching method,
exact pixel overlap of {\sc Disperse} nodes and cluster
centers, is not considered.  The clusters without {\sc Disperse} nodes
in at least 10 of the 36 matching methods and underlying 
{\sc Disperse} webs,  $\sim$ 11\%, tend to be lower in mass, have perhaps a
earlier most recent major merger, and more likely a
higher $t_2$  than the full cluster sample.  (Similar results were
found for the 1/4 clusters which didn't always have a {\sc Disperse} node
match for the combination of only the ``fixed'',''nearest'', and ``patch'' methods.)  Clusters
and their nearest {\sc Disperse} nodes within twice the smoothing length of each
other have a {\sc Disperse} node density-cluster mass relation.

\section{Cluster-cluster filaments}
\label{sec:clusclusfil}
\subsection{Assigning filaments}
The four methods above associate {\sc Disperse}
nodes with clusters for any {\sc Disperse} web.  As {\sc Disperse} nodes are linked by
{\sc Disperse} filaments, clusters matched to {\sc Disperse} nodes can
be linked to each other if their matched {\sc Disperse} nodes share a filament.  Each underlying {\sc Disperse} web and combined cluster-{\sc Disperse} node matching method produces a different set of filament assignments to cluster pairs. The cluster distribution remains fixed: these filaments are the parts of each underlying {\sc Disperse} web ``picked out'' by clusters.

In practice, filaments are assigned to cluster pairs by first
restricting the {\sc Disperse} web to {\sc Disperse} nodes which have matches to clusters. {\sc Disperse} nodes which are unmatched to clusters have their
filaments either reassigned (eventually to cluster matched {\sc Disperse} nodes) or dropped. This restriction begins by sorting unmatched {\sc
  Disperse} nodes according to the
number of filaments which end on them. Unmatched {\sc Disperse}
nodes with just one filament are simply dropped, and then a check is
done again to find whether any new unmatched {\sc Disperse} nodes with
only one filament were created, if so, these are dropped as well and
this process is repeated until no more unmatched {\sc Disperse} nodes
with just one filament remain. To drop unmatched {\sc Disperse} nodes
with more than one filament coming out, any pair of filaments meeting
at the same unmatched {\sc Disperse} node with an angle of more than
120 degrees is replaced by a single filament bypassing the unmatched
node.  After this new filament assignment, the unmatched {\sc
  Disperse} node is dropped.  If the angle is smaller, the two
filaments ending on the {\sc Disperse} node are just dropped. This is
to catch the cases where two cluster matched {\sc Disperse} nodes are
linked via an intermediate unmatched {\sc Disperse} node, where the
bending is not too large at the dropped {\sc Disperse} node. (However,
at times several of these dropped {{\sc Disperse} nodes can
  be strung together, leading to a single very bent filament.)
  Unmatched {\sc Disperse} nodes are dropped starting with those which have only two filaments
  (repeated until all instances are gone) and then going to higher multiplicities.  For
  the higher multiplicities, as dropping unmatched {\sc Disperse}
  nodes can change the number of filaments coming out of the remaining
  {\sc Disperse} nodes, unmatched {\sc Disperse} nodes which have lost
  filaments since the ordering according to filament number are passed
  over until all multiplicities are considered. At this point, all remaining
  unmatched {\sc Disperse} nodes are ordered again by
  filament number, and the process repeats until all unmatched {\sc
    Disperse} nodes are gone.

  This produces a map of cluster
  matched {\sc Disperse} nodes and their filaments.  A second map was
  also made, where only filaments directly connecting matched {\sc Disperse} nodes
  were kept, dropping the filaments interpolating through unmatched {\sc Disperse} nodes.  Unless specified
  otherwise below, the cluster-cluster filaments below also include those
  found via interpolation.  Dropping the interpolated filaments reduces the number of filaments between 8\% and 50\%, depending on web and matching variation, with the smallest smoothing and persistence having the most interpolated filaments.

The next step is to replace {\sc Disperse} nodes which are connected
by filaments with their matched
clusters.  Because the cluster to {\sc Disperse} node
matching is not always one to one, this step can be ambiguous.  A
matched node might have two clusters associated with it, or two
matched nodes might have the same nearest cluster.  (The number of
occurrences of multiple clusters within a single Hessian node patch,
used in the ``patch'' method was shown in
Fig.~\ref{fig:clusinhess}.)  And again, matching the nearest cluster
and {\sc Disperse} node within a smoothing dependent distance cut can
also lead to degeneracies because the nearest {\sc Disperse} node to a cluster might not be the nearest cluster to that {\sc Disperse} node. 
Two options are used, and appear to give similar results.  In one
case, ``nearest,'' if two clusters both have the same nearest node,
within the distance cutoff, they are linked to each other with
filaments, and if two nodes claim the same cluster as their nearest
cluster, and another cluster claims either node, the two clusters are
also linked via filaments. A second way to proceed, ``fixed dist,'' is
to match every cluster within the smoothing dependent distance cut to
the node, and if there is more than one, to connect these clusters
to each other with filaments.

At the end of this construction,
every cluster has a list of other clusters to which it is connected
via filaments.  There is a different set of cluster-cluster filament pairs for each combination of cluster-{\sc Disperse} node matching
method and underlying original {\sc Disperse} web.

\begin{figure*}
\begin{center}
\resizebox{3.2in}{!}{\includegraphics{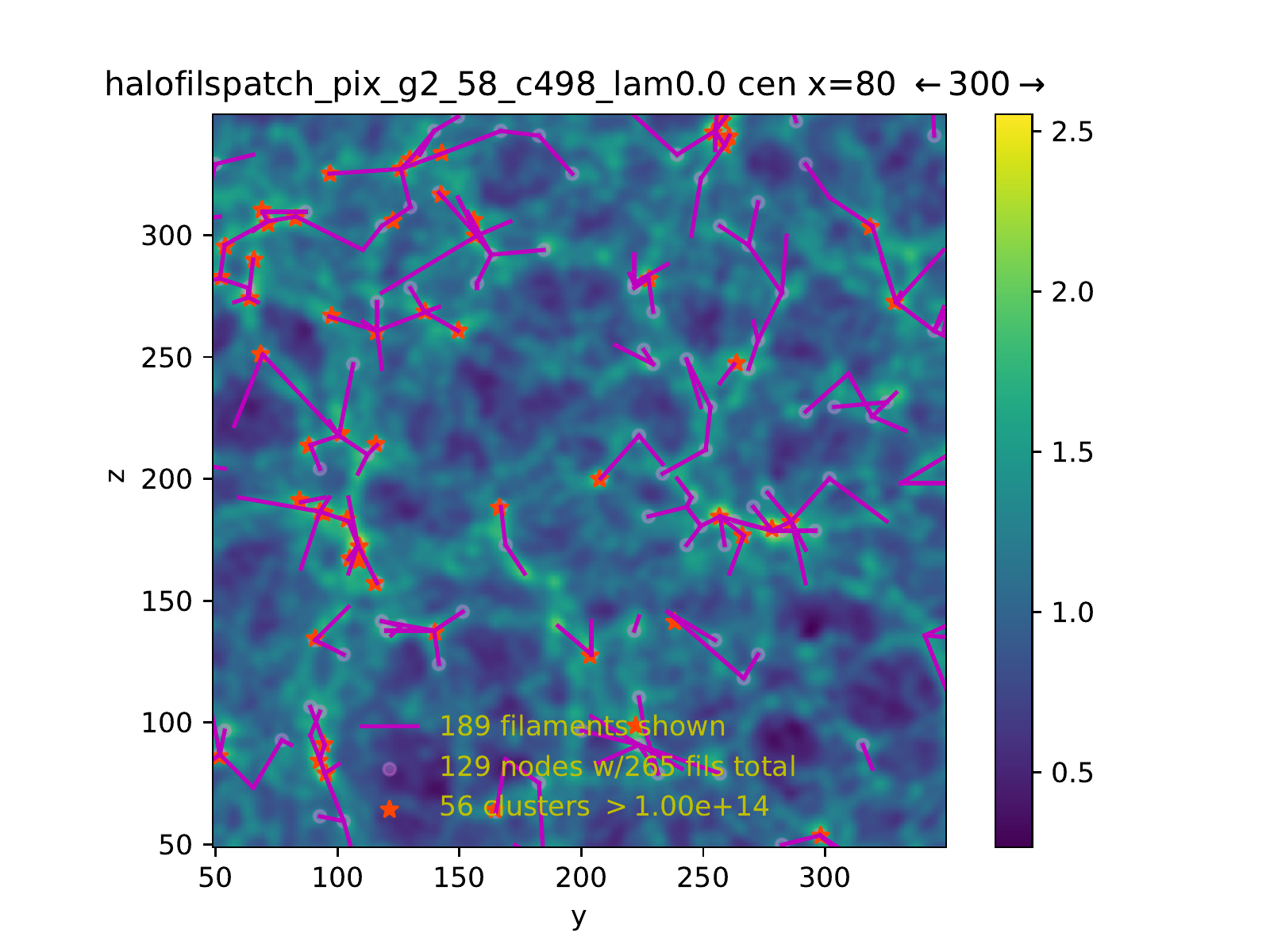}}
\resizebox{3.2in}{!}{\includegraphics{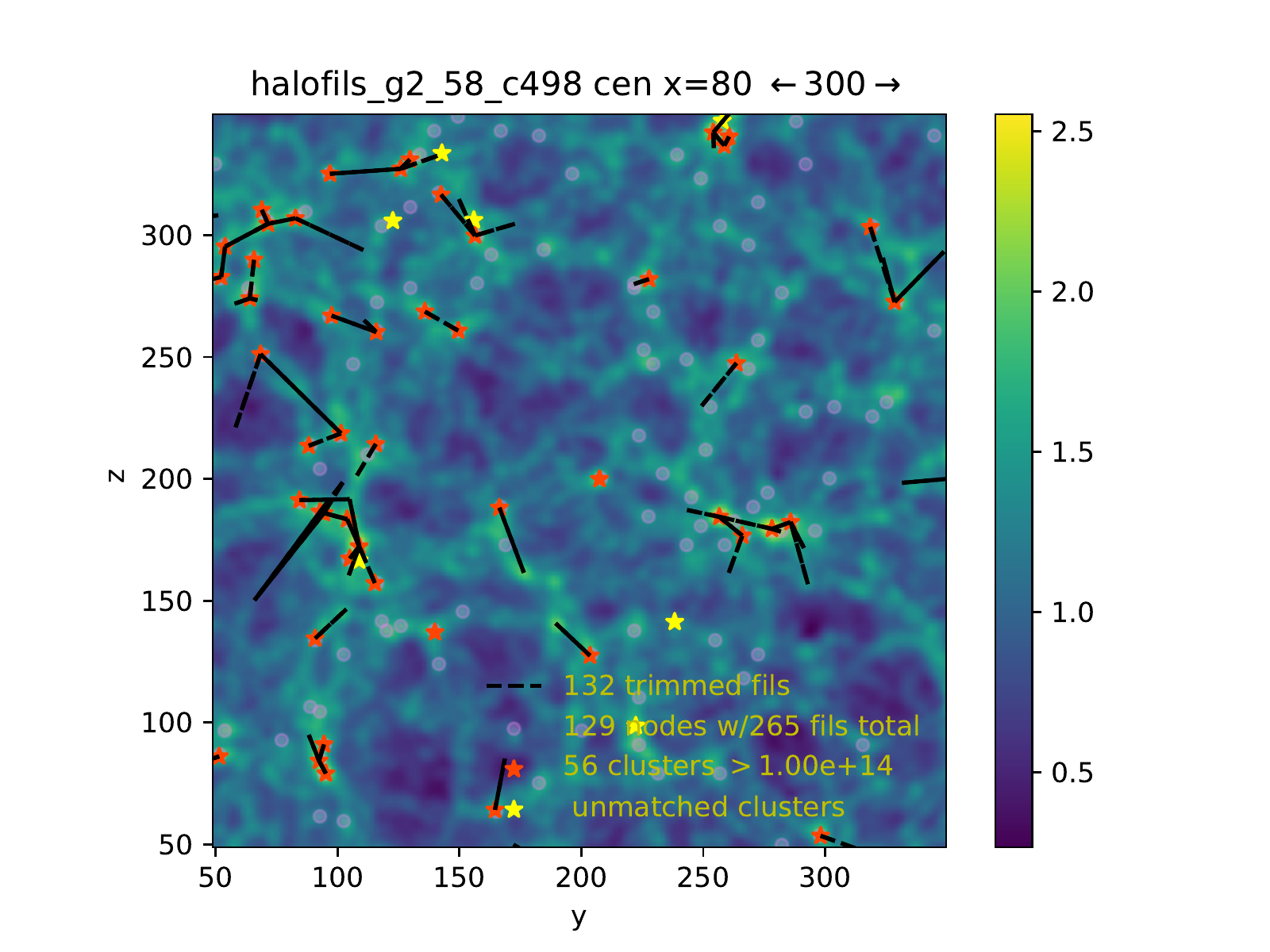}}
\resizebox{3.2in}{!}{\includegraphics{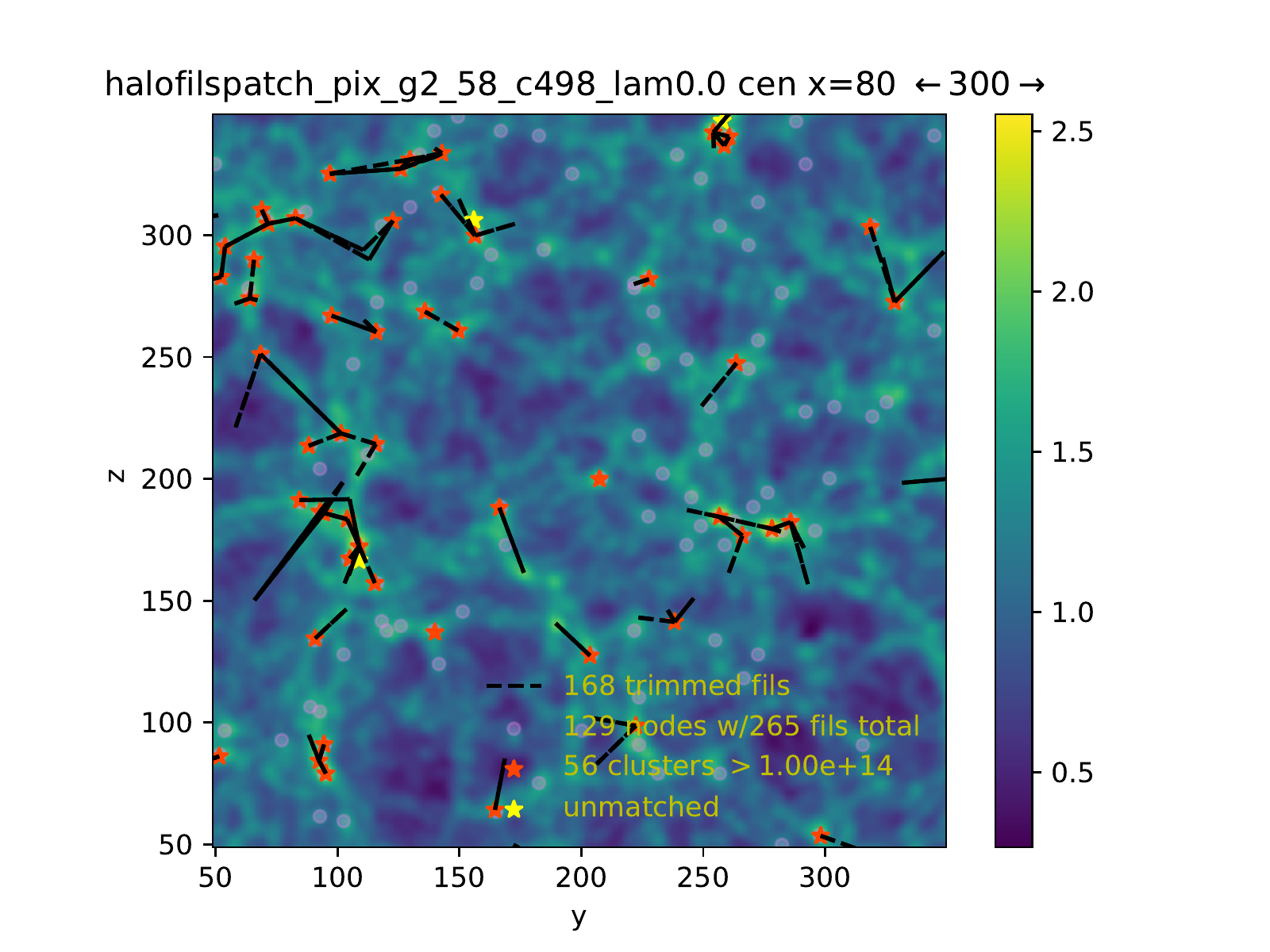}}
\resizebox{3.2in}{!}{\includegraphics{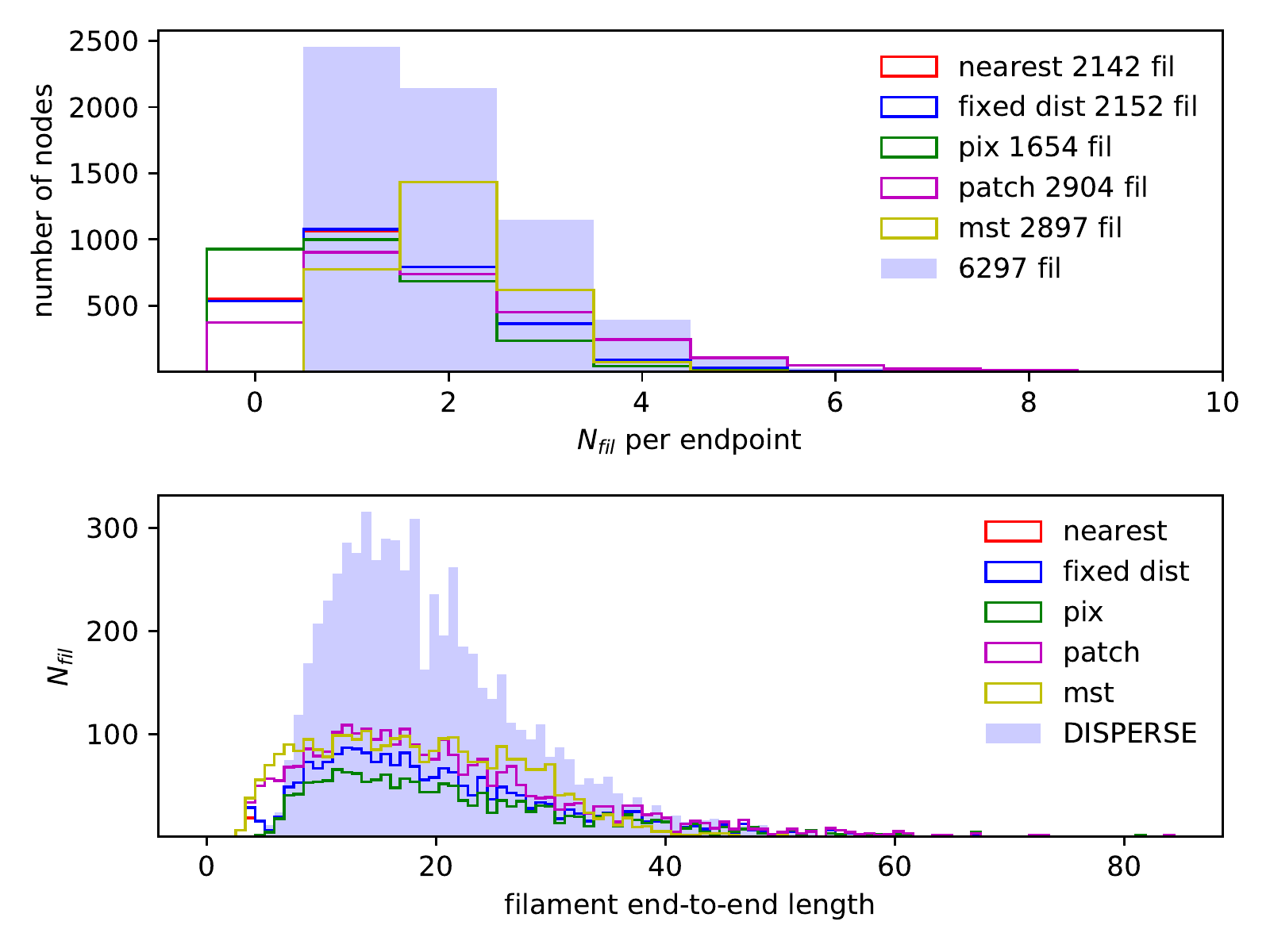}}
\end{center}
\caption{Comparisons of a 30 $Mpc/h$ slice for filaments resulting from the 3$\sigma$ persistence, 2 $Mpc/h$ smoothed {\sc Disperse} web, 
  overlaying a (logarithmic) density map (from Fig.~\ref{fig:dispall3}).  
Interpolation through unmatched {\sc Disperse} nodes is done as described in the text.  At upper left is the
  original {\sc Disperse} web.  The (magenta) {\sc Disperse} filament endpoints are {\sc
    Disperse} node pixels.  The red stars are clusters.  For the cluster-cluster filaments,
  unmatched clusters are yellow stars, and the black lines show cluster pairs which are
  connected by using their matched {\sc Disperse} nodes and either the
  ``nearest'' method (upper right) or ``patch'' method (lower left),
  described in \S \ref{sec:clus2node}. Log scale is used for
  density. At lower right are statistics for all the different matching methods
  described in \S\ref{sec:clus2node} plus the cluster based
  minimal spanning tree (``mst'') described in the text.  The top plot in this panel is the number of
  filaments per cluster(with total filaments per method in the
  legend), the bottom plot is the filament length distribution.  All maps
  have the same clusters and original fixed underlying {\sc Disperse} web (its statistics are shown as
  shaded bars, repeated from Fig.~\ref{fig:dispall3}).  Clusters can
  have no filaments due to being unmatched to a {\sc Disperse} node or
  by having their matched {\sc Disperse} node not connected by
  filaments to other cluster-matched {\sc Disperse} nodes (either directly or via dropped {\sc Disperse} nodes).}
\label{fig:inducedfilaments}
\end{figure*}

\subsection{Comparing different cluster-cluster filament assignments}
\label{sec:inducedfilaments}

 An underlying {\sc Disperse} web is shown, along with
 two cluster pair filament assignments based upon it, in Fig.~\ref{fig:inducedfilaments}.
As in Fig.~\ref{fig:dispall3} this is for a 30 $Mpc/h$ density slice, with the original {\sc
  Disperse} web from Fig.~\ref{fig:dispall3} at upper left (magenta filaments, smoothing 2 $Mpc/h$
with $3\sigma$ persistence), the filaments produced from the ``nearest'' method
at upper right and the filaments produced by the ``patch'' method at lower left (cluster
{\sc Disperse} node matching
methods are described in \S\ref{sec:clus2node}).  The matched clusters are
red stars, unmatched clusters are yellow stars, and the cluster-cluster
filaments are shown as black lines.  At lower right are the statistics of
all the filament assignments based upon this particular {\sc
  Disperse} smoothing and persistence.  

Included in the comparison of cluster-cluster filament pairs in
Fig.~\ref{fig:inducedfilaments} (lower right panel) is a cluster
based minimal spanning tree web (some variants are in, e.g.,
\citet{BarBhaSon85,ParLee09,Alp14,Per20}). This is included for
comparison because it is a web constructed
directly from the clusters themselves, by choosing each cluster as a
web node.  It is created by ranking cluster pairs according to
some property (two ranking properties which have been used for halo/galaxy based webs are the distance
between them, \citet{Alp14}, used here, or $M_1 M_2/r^2$
\citet{Per20}).  Filaments are then assigned to cluster pairs in
ranked order, omitting any pair where both proposed endpoints are
already connected to filaments, to get $N_{clus}-1$ filaments.

In the filament assignments for the example shown in Fig.~\ref{fig:inducedfilaments}, with an underlying 2 $Mpc/h$ smoothing, 3$\sigma$ persistence {\sc Disperse} web, some general trends can be seen.
As the cluster-cluster filament pairs only have 2898
clusters to serve as possible endpoints, versus the 6295 original
{\sc Disperse} web nodes, there are fewer cluster-cluster
filaments (filament counts are listed in the legend of
Fig.~\ref{fig:inducedfilaments}, top half of lower right panel). There are also some longer cluster-cluster filaments relative to those of the
underlying {\sc Disperse} web, likely due to the merging of shorter
filaments when interpolating between cluster matched {\sc Disperse} nodes.
The filaments
tend to be shorter in the cluster based minimal spanning tree web, as the shortest
distance pairs were chosen to have filaments.

More generally, in the original {\sc Disperse} webs, and the minimal spanning tree webs,
every cluster has at least one filament (although the minimal spanning
tree web tends to be more tree than ``web'' like, as it does not have
closed loops). The cluster-cluster filaments do not result in a fully connected object, in particular, there are 
clusters with no filaments, either because they
have no associated {\sc Disperse} nodes (see Fig.~\ref{fig:halonodepix}) or because
their matched {\sc Disperse} node didn't have filaments to another
matched {\sc Disperse} node.  (However, some unmatched clusters acquire filaments in the ``fixed'' method for assigning filaments, even though they don't directly have associated {\sc Disperse} nodes.) Considering all the cluster-cluster filament pairs, based upon all the different {\sc Disperse} webs, the fraction of
clusters with no filaments (for either of these reasons) is highest for
pixel matching (``pix''), and lowest for 
Hessian node patches (``patch'').  For ``patch'', ``nearest,'',
``fixed dist,'' for smoothing below 5 $Mpc/h$, the fraction of
clusters with no filaments lies between $\sim$ 10\% -$\sim$25\%, while for ``pix'' the number of unlinked clusters was
above 30\% for all smoothings.  The 5 $Mpc/h$ smoothing reached 95\%
unlinked clusters (for ``pix'') and more generally was worse for
linking clusters for all methods.  
In the 36 cluster-web matching combinations with smoothing $< 5 Mpc/h$, 47 of the 2898 clusters
are never connected via a filament to another cluster. 
If filaments aren't interpolated across unmatched {\sc Disperse} nodes, this goes up to 152 clusters. 
\begin{figure}
\begin{center}
\resizebox{3.5in}{!}{\includegraphics{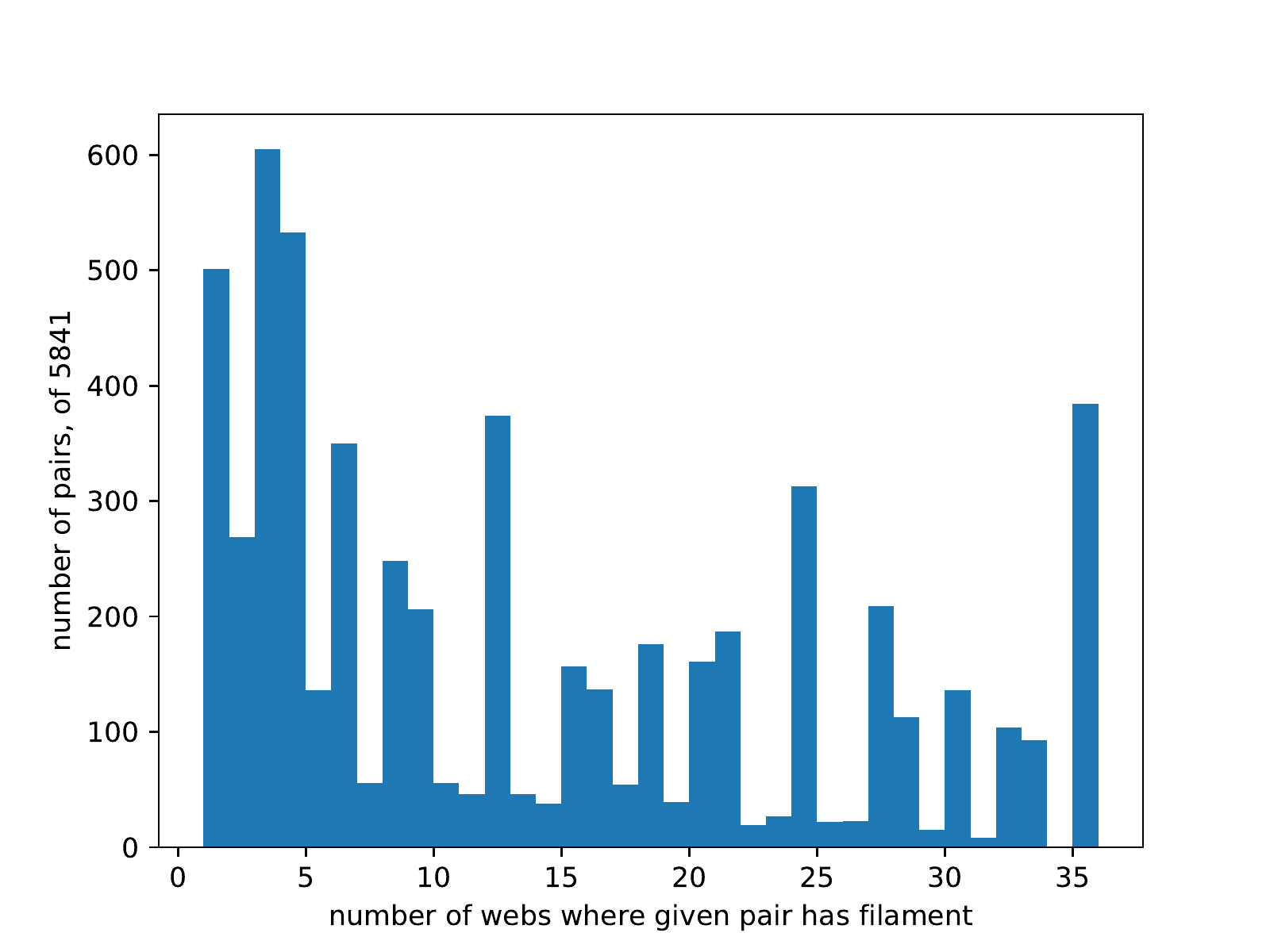}}
\end{center}
\caption{Frequency of cluster pairs with filaments.  The number of times a
  given cluster pair has a filament, out of the 36 persistence/smoothing/matching combinations; 383 pairs
  had filaments in all 36.  Only the
  5841 cluster pairs linked with a filament in at least one of the 36
  cluster- web matching combinations are shown.}
\label{fig:numtimespairs}
\end{figure}


\begin{figure}
\begin{center}
\resizebox{3.5in}{!}{\includegraphics{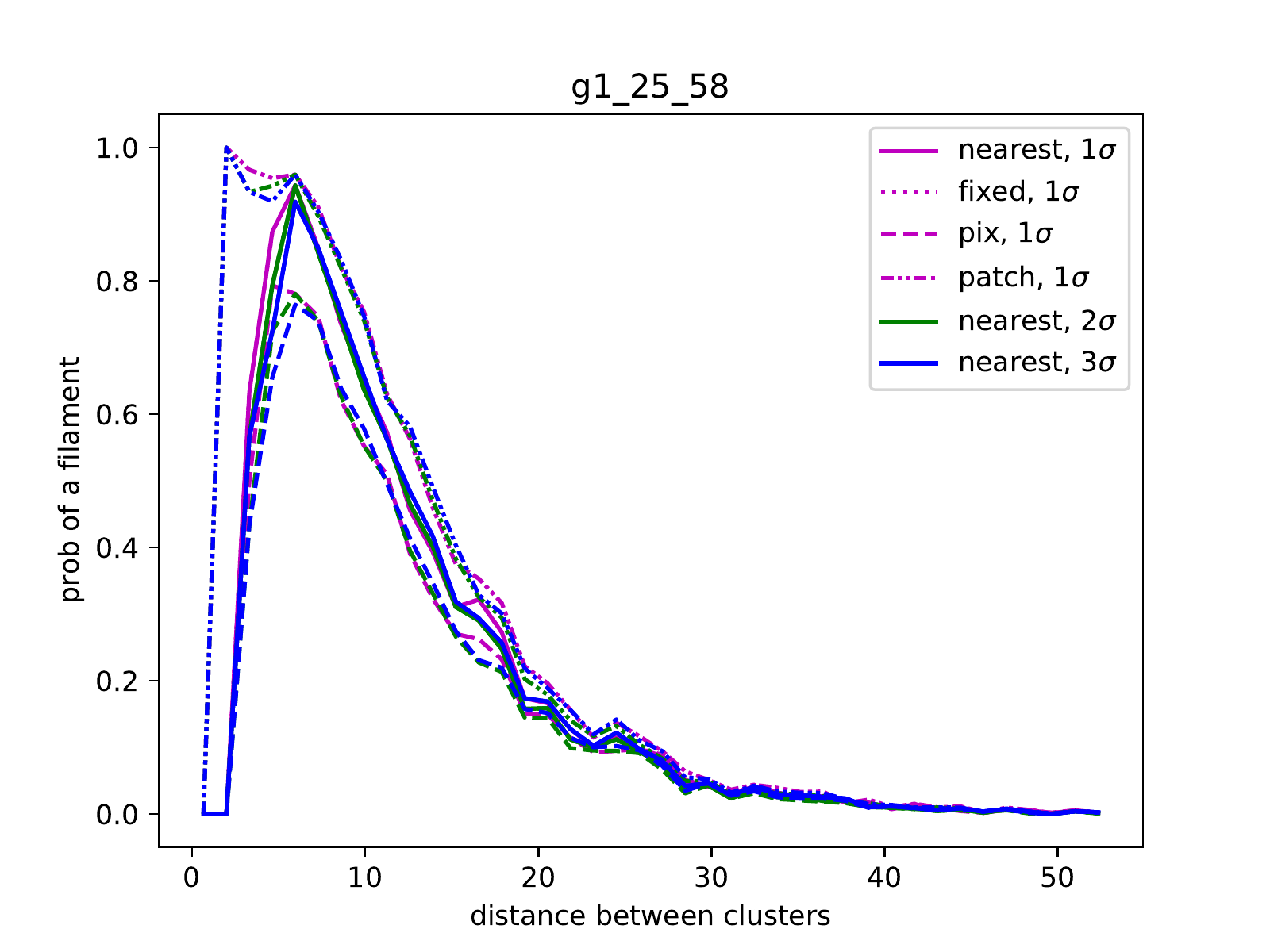}}
\resizebox{3.5in}{!}{\includegraphics{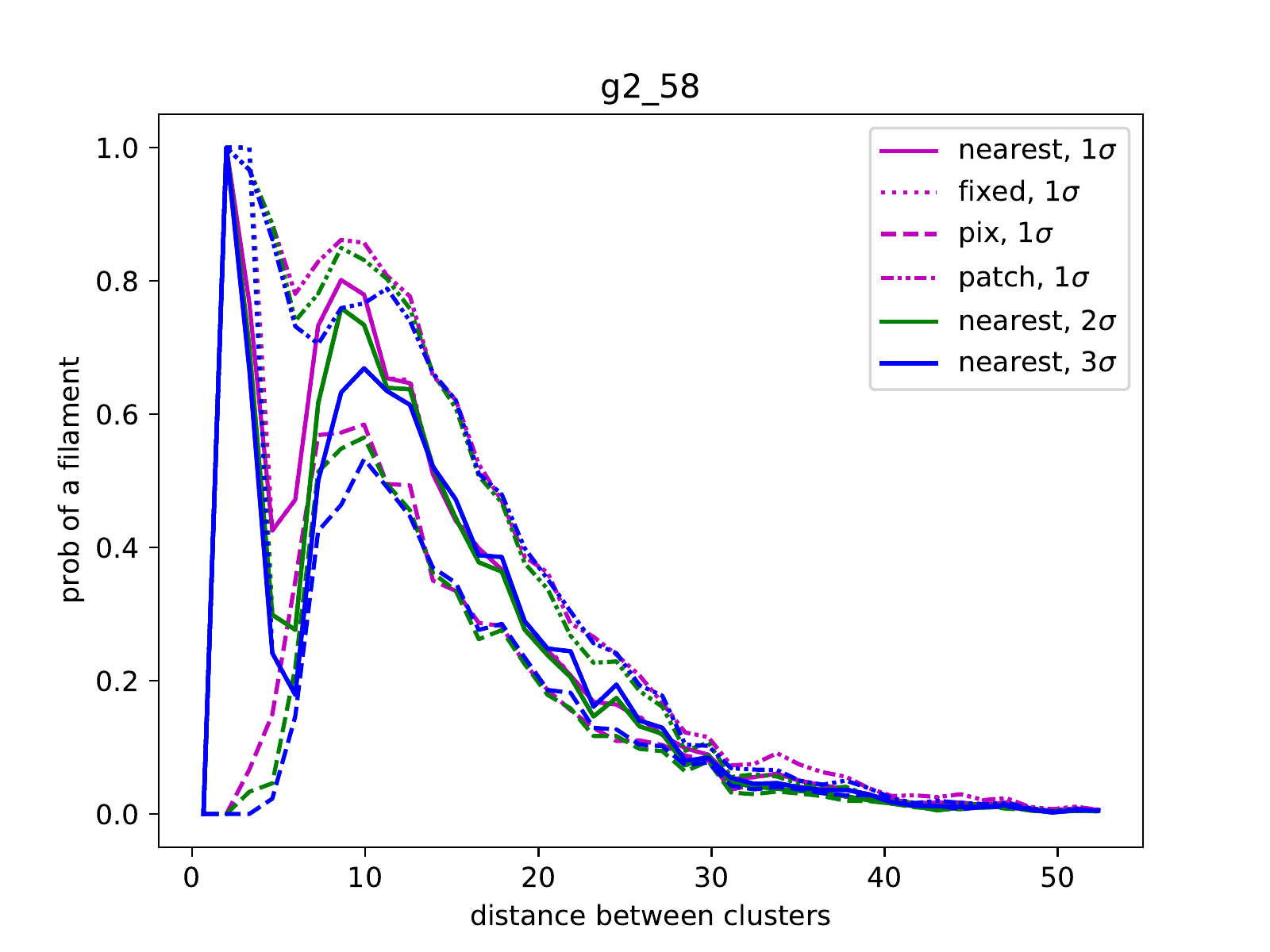}}
\resizebox{3.5in}{!}{\includegraphics{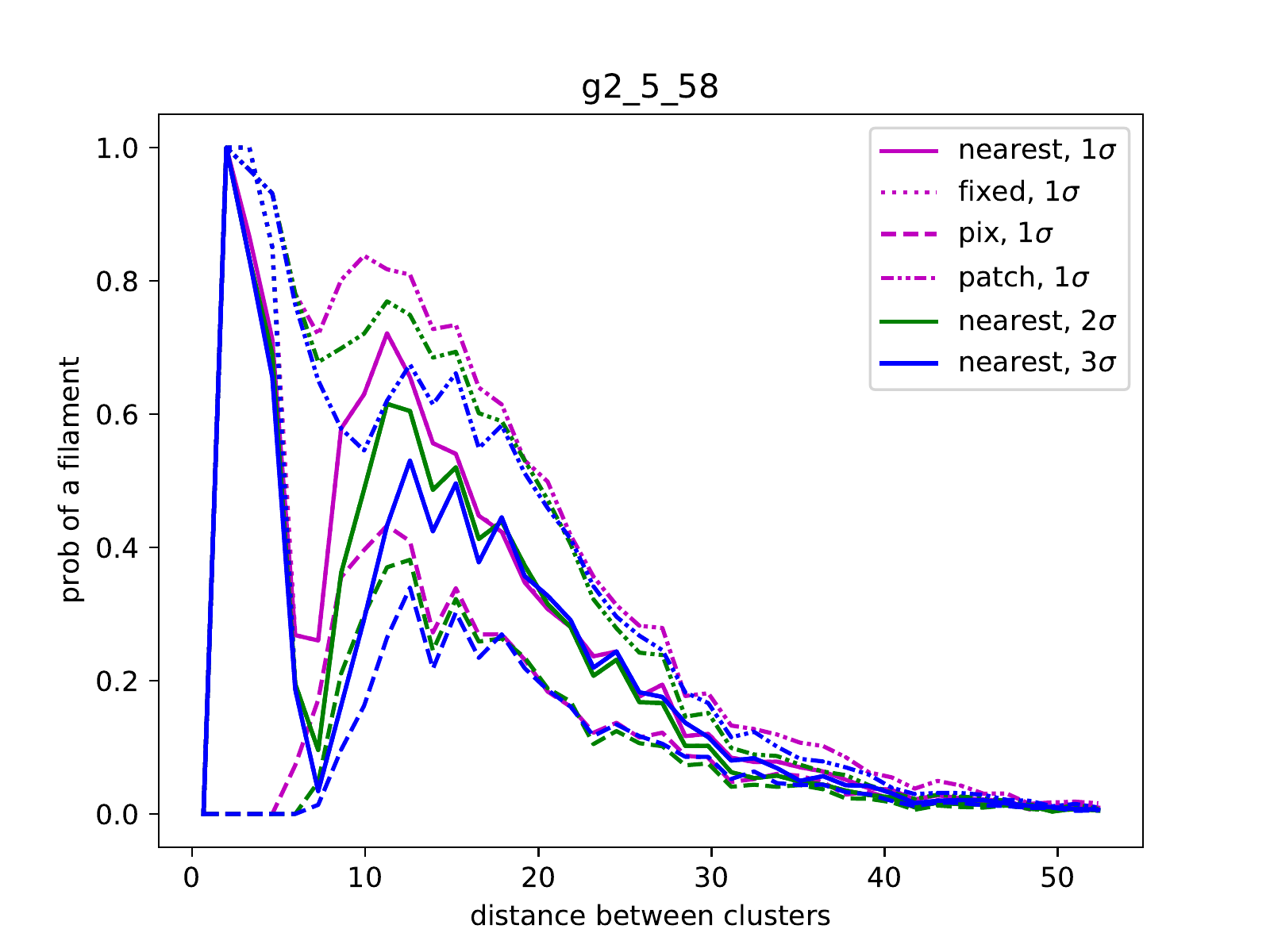}}
\end{center}
\caption{Fraction of cluster pairs with a filament  as a function of
  cluster separation. The panels, top to bottom, correspond to
  smoothing lengths (1.25 $Mpc/h$, 2 $Mpc/h$, 2.5 $Mpc/h$). Line
  types distinguish cluster-{\sc Disperse} node correspondence, color
  denotes {\sc Disperse} persistence chosen.  The Hessian patch
  matching between clusters and the {\sc Disperse} web gives the
  simplest relation between filament likelihood and cluster
  separation, roughly less likely as separation increases, although for all webs based on smoothing above 1.25
  $Mpc/h$ there are additional features as a function of cluster
  separation.}
\label{fig:probfil}
\end{figure}

\begin{figure}
\begin{center}
\resizebox{3.5in}{!}{\includegraphics{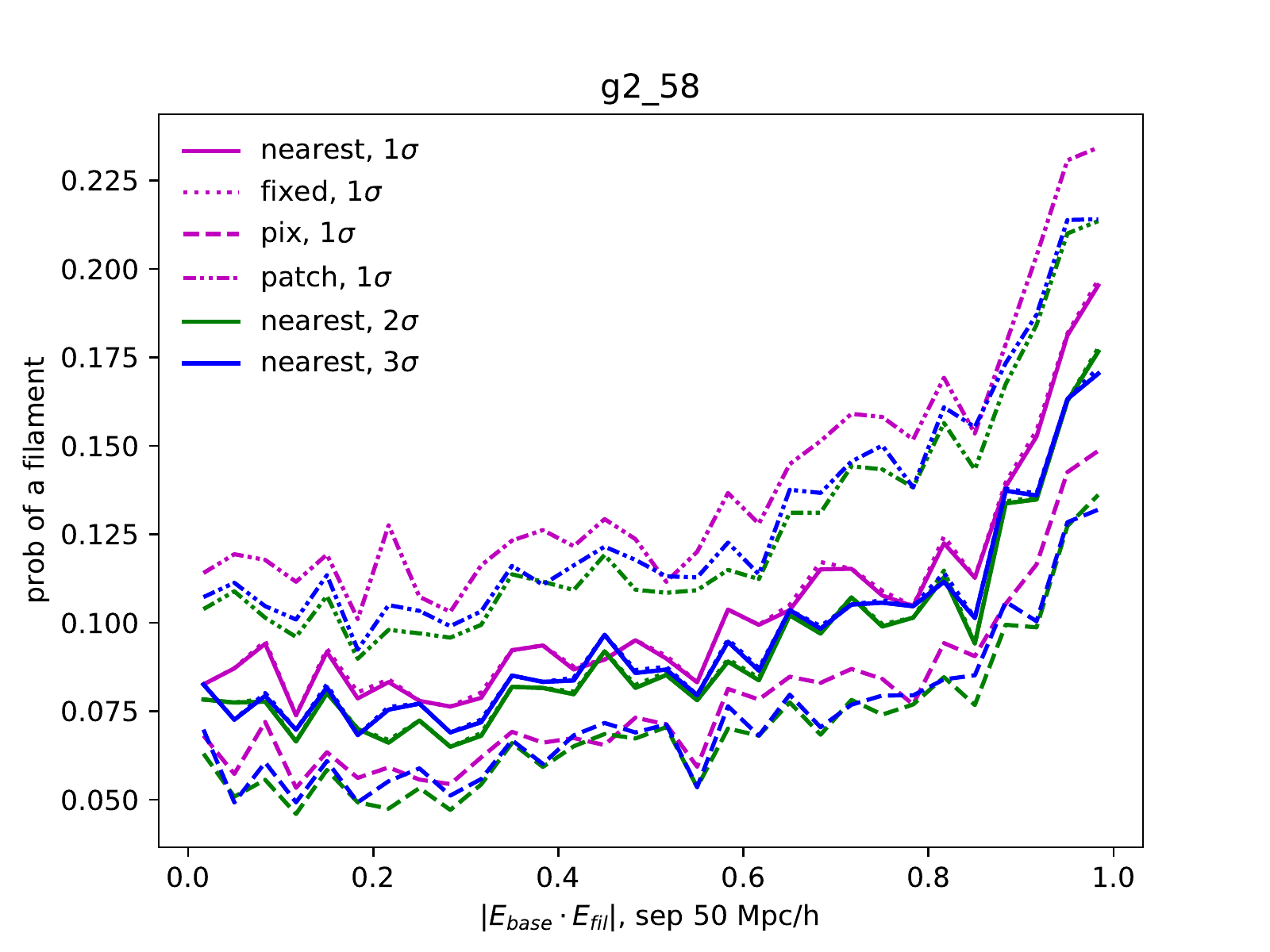}}
\resizebox{3.5in}{!}{\includegraphics{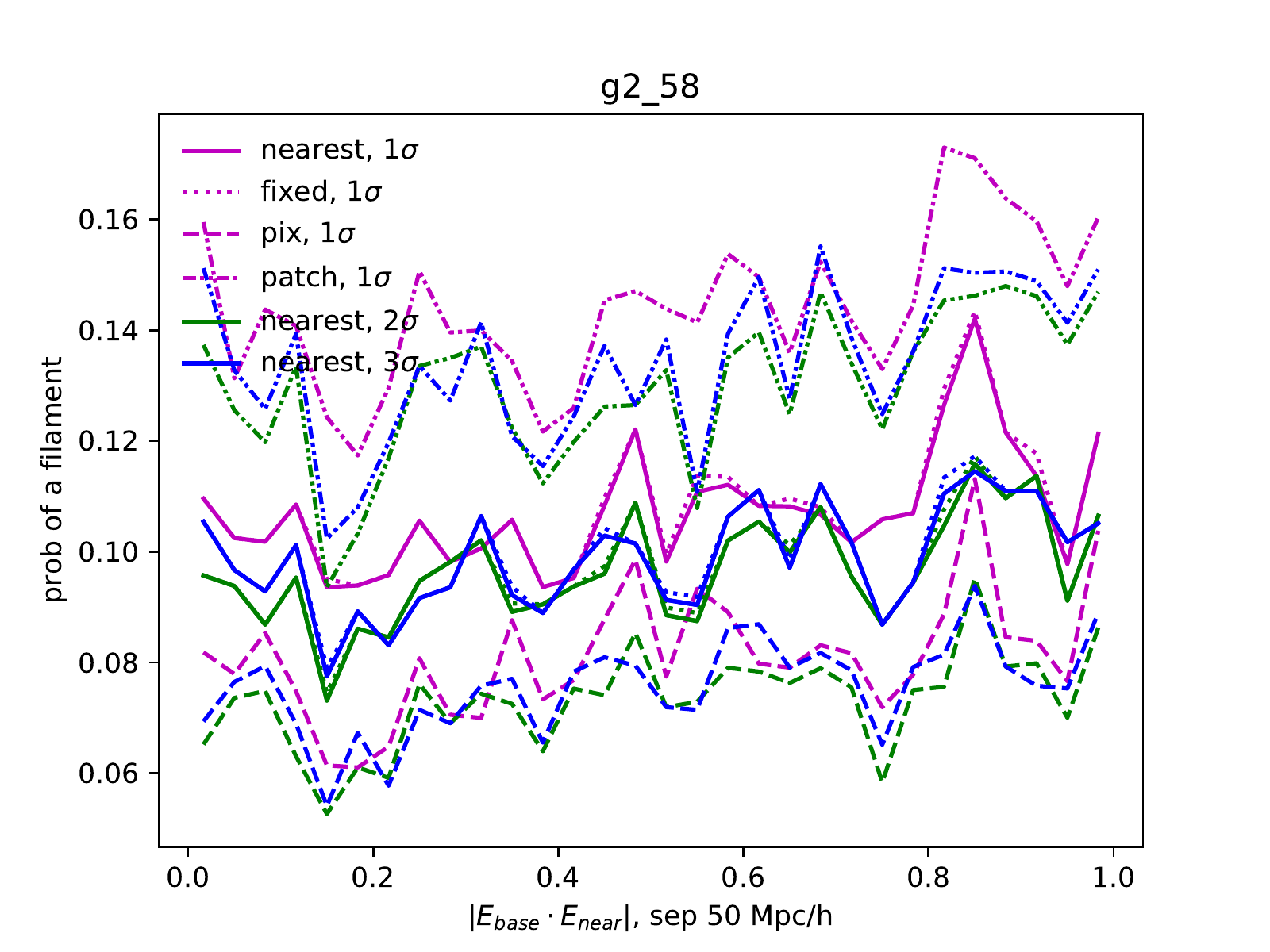}}
\end{center}
\caption{Fraction of cluster pairs separated by less than $\sim$50
  $Mpc/h$ which have a filament, for different cluster to node
  matching methods as described in \S\ref{sec:clus2node}.  Top: as a
  function of alignment of a cluster axis and cluster-cluster pair
  axis. Bottom: as a function of alignment of the two cluster pair
  endpoints.  Smoothing length for both is 2 $Mpc/h$.  Again, line types
  distinguish cluster-{\sc Disperse} node correspondences, color
  denotes chosen {\sc Disperse} persistence.}
\label{fig:dotfit}
\end{figure}
\section{Cluster pairs}
\label{sec:cluspair}

\subsection{Cluster pairs with and without filaments}

All of these sets of inherited cluster-cluster filaments, from each of
the 36 {\sc Disperse} web-matching combinations, start with the same set of clusters.
One way of
thinking about the cluster-cluster filaments which appear, based upon a
particular matching method and underlying {\sc Disperse} web, is as an operation on the
distribution of cluster pairs which selects those pairs which are linked.
(Given the huge numbers of clusters with no filaments when 5 $Mpc/h$ smoothing is used, these {\sc  Disperse} webs and their associated cluster-cluster filament pairs will not be discussed in this section.)

Of the over 420,000 cluster pairs in the box, only 5841 pairs have a filament
in any of the 36 web variations. If interpolation is dropped, that is,
only underlying {\sc Disperse} web filaments with both ends lying on
cluster matched {\sc Disperse} nodes are kept, only 4440 cluster
pairs ever have filaments.  The distribution of how often each pair
appears in these 36 web variations is shown in Fig.~\ref{fig:numtimespairs}.  
Of these, 383 cluster pairs have filaments in every web, and these tend to be at smaller separation, e.g., between 10-20 $Mpc/h$, although one cluster
pair which always has a filament is separated by over 65 $Mpc/h$.
Slightly fewer than 1/5 of the total number of pairs which were ever linked 
have filaments in most (i.e., at least 3/4 of the 36
web persistence/smoothing/matching combinations; a bit more than half of the
1/5 frequently linked cluster pairs correspond to filaments directly, not interpolated,
between two cluster matched {\sc Disperse} nodes).  About 1/3 of
the cluster pairs ever having filaments only have them rarely (for
$\leq 4$ of the 36 combinations), while almost half have filaments only for $\leq 9$ of the 36 persistence/smoothing/matching combinations.
Thus, the 36 ways of assigning filaments to cluster pairs
produce 36 sets of mostly non-coinciding cluster pairs with filaments. The rarest pairs seem to occur for ``patch''
matching, and for the largest smoothing, likely because ``patch'' can
match many clusters across a large region with the same {\sc
  Disperse} node or nodes.

Some trends appear in the pairs of clusters which are linked by filaments.  As might be expected, clusters which are closer together are more
likely to have a filament between them.  In more
detail, for these 36 webs, the fractions of cluster pairs which are
connected by a filament, as a function of separation, are shown in
Fig.~\ref{fig:probfil}.  A dip in probability for somewhat close
pairs is noticeable for the ``nearest'' and ``fixed dist'' filament
assignments to cluster  pairs, for smoothings 2, 2.5$Mpc/h$, although it is slightly present in the ``patch'' web as well.\footnote{As
clusters matched to the same {\sc Disperse} node are linked by
filaments, increasing the radius within which clusters and {\sc
  Disperse} nodes are matched would remove the drop in number of
linked cluster pairs, however, this dip is at the scale where nearby
clusters tend to appear, so changing the matching radius would also tend to match more
cluster pairs to a single {\sc Disperse} node. Another possibility is
to add yet another parameter in creating the filament assignments for
``nearest'' and ''fixed'', a separate minimum radius within which all
clusters are linked.}    For a given separation bin, cluster pairs with filaments were a minority of
cluster pairs beyond 15-20 $Mpc/h$ separations for any given separation.

Pair separation is the strongest indicator of whether a cluster pair
is likely to be linked (but not sufficient information, as seen in
Fig.~\ref{fig:probfil}).  Alignments of the clusters' long axes are
another property that is expected to make a filament more likely
\citep{BonKofPog96,BonMye96}.  Taking pairs with separation less than
$\sim$ 50 $Mpc/h$, 
the fraction of linked pairs did
not seem to have a strong dependence upon cluster-cluster axes
alignment (e.g., bottom of Fig.~\ref{fig:dotfit}, which for 2 $Mpc/h$
smoothing shows only perhaps a slight increase, and seems slightly stronger
for smaller smoothing).  However, when the cluster-cluster pair axis was
aligned with either cluster axis, a filament was present around twice as
often as when these two axes were perpendicular to each other (although
still only $\leq$ 25\% of the time, largest for the largest smoothing), some examples are shown in
Fig.~\ref{fig:dotfit}, at top.  

Instead of looking at pair features one-by-one to see if a filament is present, machine learning can try to
answer the classification problem (given a pair, is there a filament present or not)
using a combination of features.  Some methods also assign ``importance'' of each
feature to the classification.  Using the Random Forest classifier,
``out of the box''\footnote{RandomForestClassifier(max\_depth=5,
  n\_estimators=10, max\_features=1), imported
  from sklearn.ensemble, described in {\tt
    https://scikit-learn.org/stable/modules/ensemble.html} }, the
relative importance of the features of distance, cluster-pair axis alignment,
cluster-cluster alignment, and rank, i.e. how many clusters were
closer to the endpoint cluster than the other endpoint of the pair,
were considered for all pairs below a fixed separation $\sim 50 Mpc/h$. 
Pair separation was, as expected, the most ``important'' feature,
followed by rank (highly correlated with pair
separation). When considered along with pair
separation, cluster-cluster alignment and cluster-pair axis
alignment were both approximately of the same ``importance,'' in
contrast to the differing dependence seen when these were considered
on their own in Fig.~\ref{fig:dotfit}.  It is possible that
considering cluster alignment separately from cluster separation, as
in Fig.~\ref{fig:dotfit},
weakened the signal.  
It is also true, however, that for many cases, the machine learning success of matching
filaments to pairs was not that good (above 8\% mismatch for 11/36 of
the filament assignments to cluster pairs), so it might be that more tuning is needed to use this approach,
or that the the presence of a filament is not amenable to being
predicted by these (few) parameters. The AdaBoost classifier was also
tried\footnote{AdaBoostClassifier(), also from sklearn.ensemble}, but
did quite poorly on some of the 36 variations of cluster-cluster filament pair assignments.

Conversely, {\sc Disperse} filaments with higher density at their critical points are also more likely to have matched cluster pairs, perhaps in part because they are the ``low'' points relative to higher density nodes. 
The analogous plots to Fig.~\ref{fig:probfil} for filaments (what fraction of filaments of a given length are associated with a cluster pair) show that short distance filaments ($<$5-10 $Mpc/h$) are more likely to have pairs.  However the fraction of filament pairs matched to cluster pairs then only decreases slowly out to 40-50 $Mpc/h$, unlike the sharp decrease in the fraction of matched cluster pairs with increased separation in Fig.~\ref{fig:probfil}.   

\begin{figure}
\begin{center}
  \resizebox{3.5in}{!}{\includegraphics{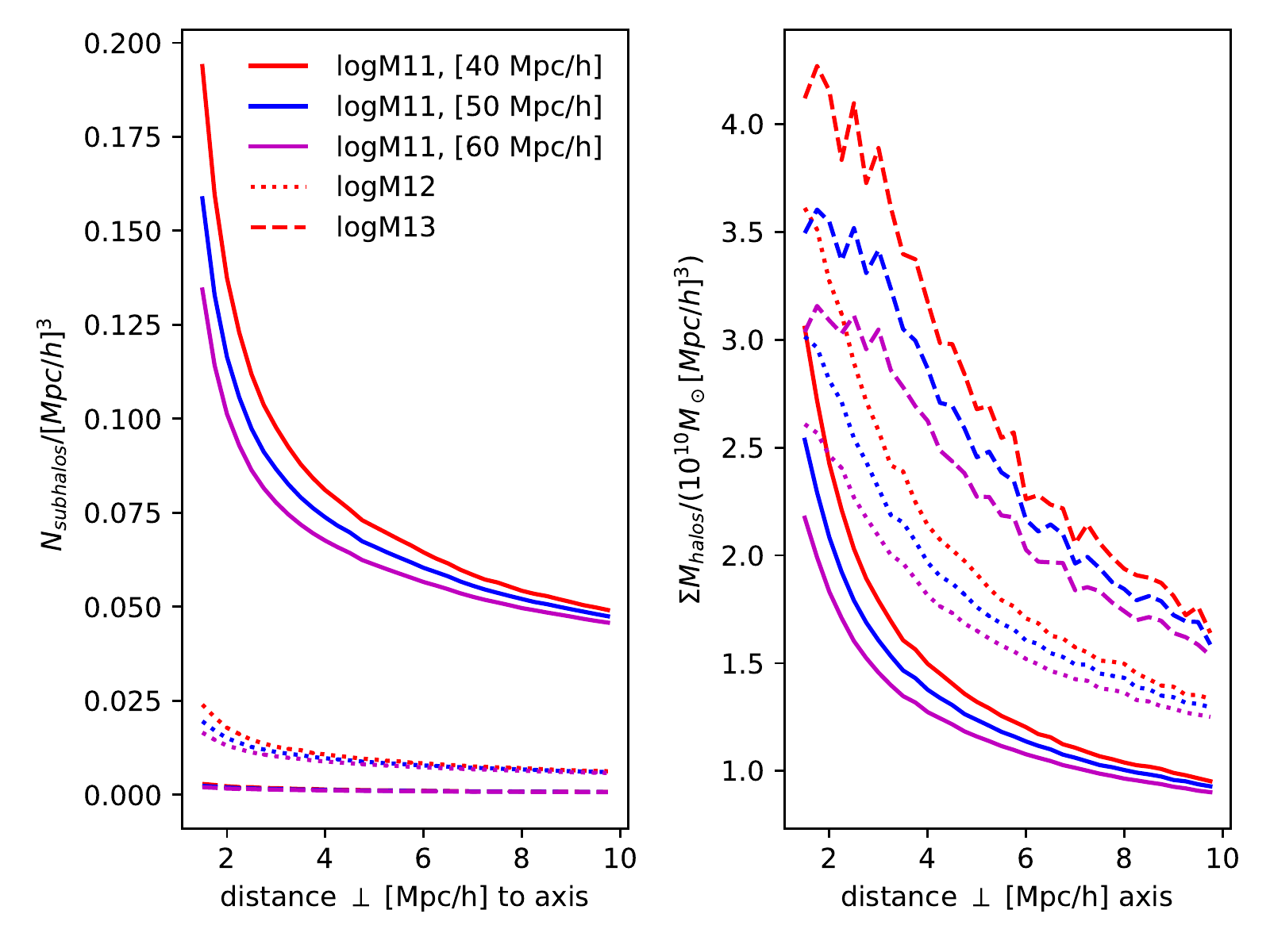}}
  \resizebox{3.5in}{!}{\includegraphics{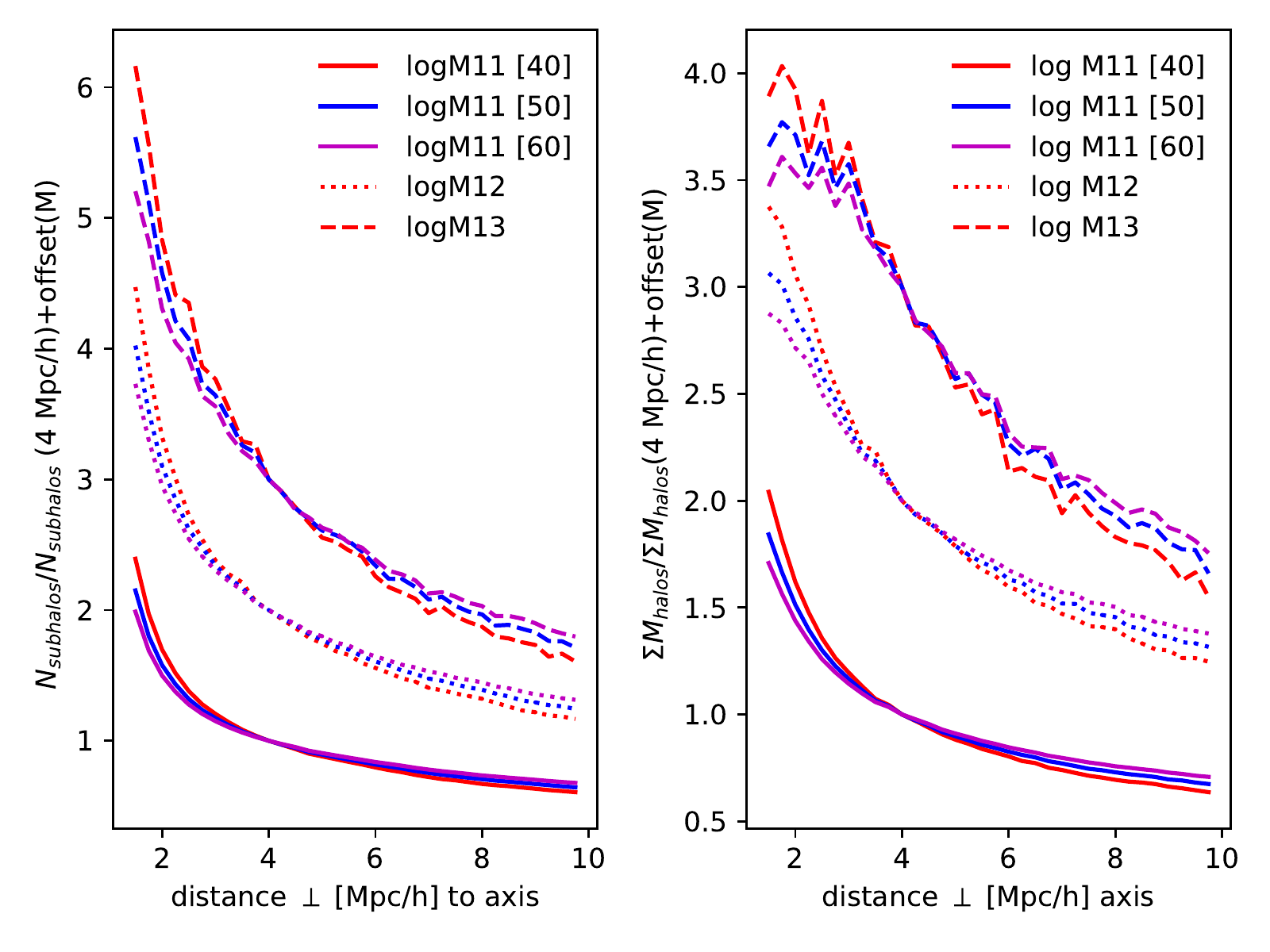}}
\end{center}
\caption{Enhanced density of subhalo counts (left) or halo mass
  (right), between cluster pairs, for (sub)halo masses  $10^{11}-10^{12} M_\odot$
  (solid), $10^{12}-10^{13} M_\odot$ (dotted) and $10^{13}-10^{14} M_\odot$
  (dashed). Top: the radial subhalo number density (left) and mass
  density (right) perpendicular to the line linking all pairs of
  neighboring clusters separated by $\leq 40 Mpc/h$ (red), $50 Mpc/h$
  (blue), and $60 Mpc/h$ (magenta), with 11996, 20906, and 33842
  cluster pairs respectively.  The halo masses are placed at the halo center (treated as point masses).
  Below, counts and mass density rescaled by value at 4 $Mpc/h$, and then multiplied by $\log_{10}
  M_{min}/10^{10} M_\odot$, to set apart the different mass
  populations.  The minimum distance from the cluster-cluster axis is taken to be 1.5 $Mpc/h$ to avoid numerical issues.}
\label{fig:pairprof}
\end{figure}


\begin{figure}
\begin{center}
  \resizebox{3.5in}{!}{\includegraphics{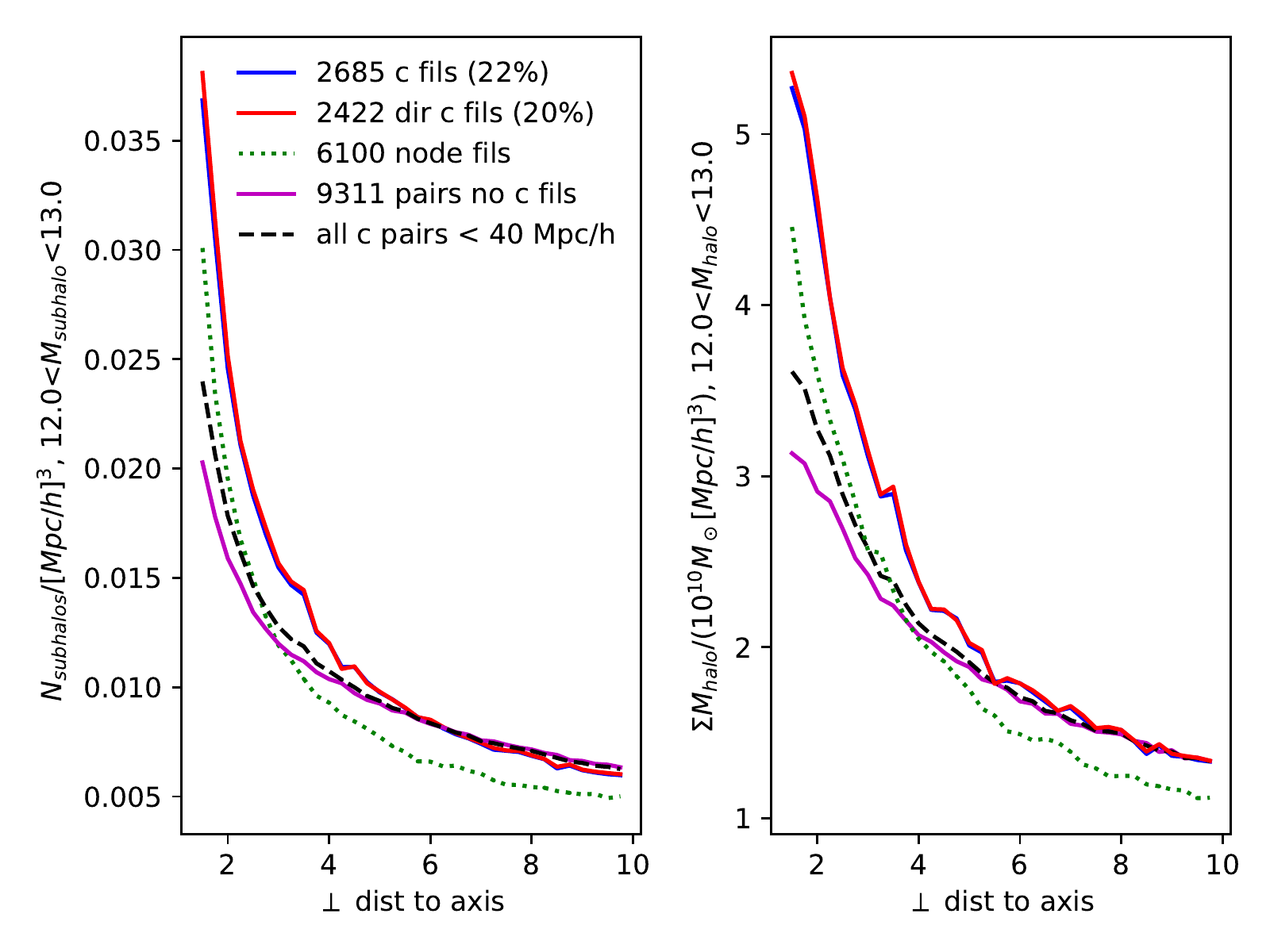}}
\end{center}
\caption{Example of the radial subhalo number count density (left) and halo 
  mass density (right), for subhalo mass or halo mass respectively in the
  range $10^{12}
  M_\odot-10^{13} M_\odot$, perpendicular to the line connecting cluster or {\sc
    Disperse} node pairs with separation
  $\leq$ 40 $Mpc/h$.  The different pairs are clusters connected by
  filaments either directly (``dir c'', red solid line) or also via
  interpolation (``c'', blue solid line), not connected by filaments
  (bottom magenta line), all cluster pairs (black dashed,
  same curve as in Fig.~\ref{fig:pairprof}) and {\sc Disperse} nodes
  connected by filaments (dotted line). The underlying {\sc
    Disperse} web was for 2 $Mpc/h$ smoothing and  $3\sigma$
  persistence, with matching to clusters via the ``patch'' method.
  The filament profiles are closer to $1/r$ than the $1/r^2$ seen for other definitions by \citet{ColKruCon05,AravanJon10}.}
\label{fig:pairfilprof}
\end{figure}
\begin{figure}
\begin{center}
  \resizebox{3.5in}{!}{\includegraphics{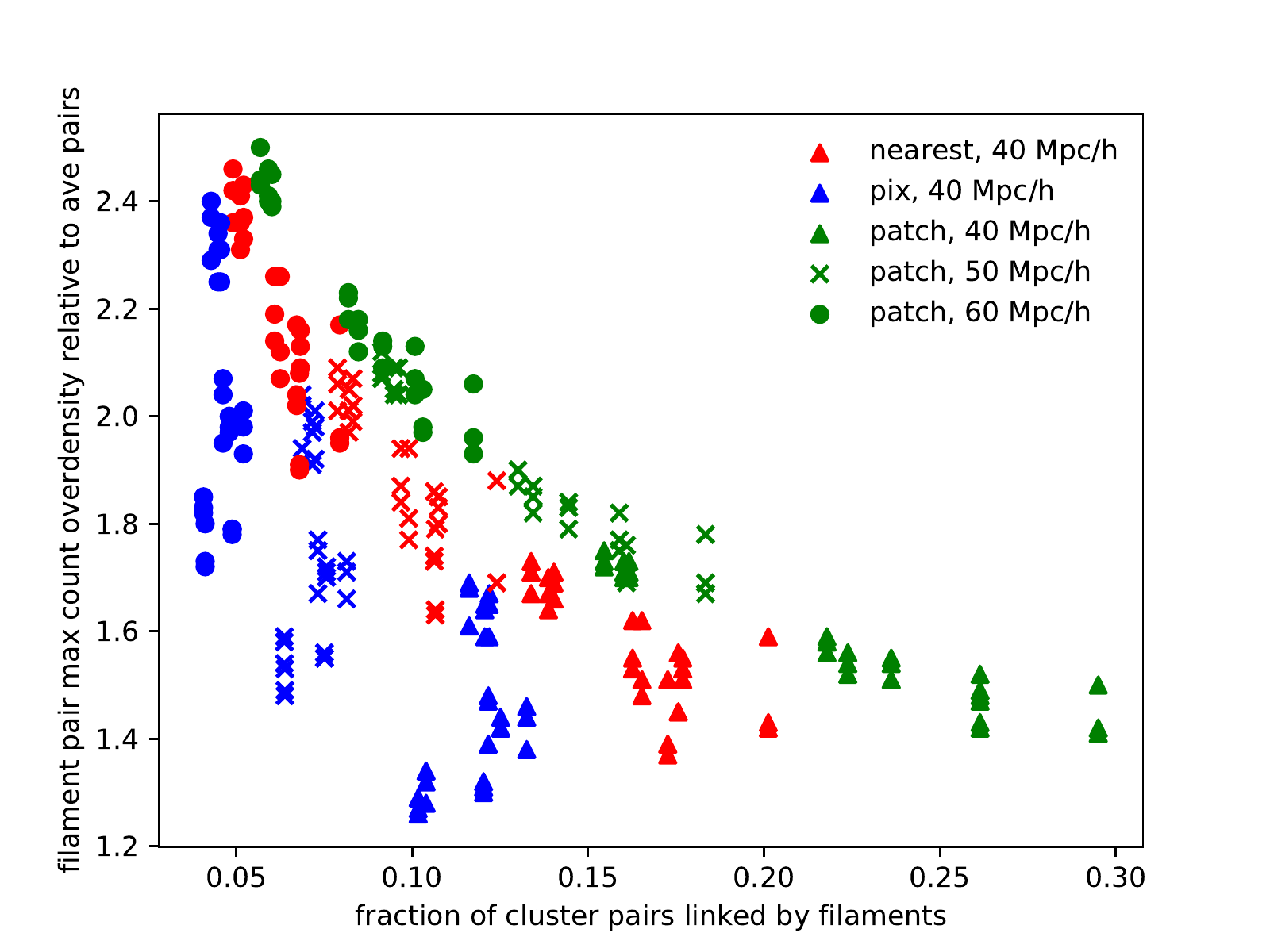}}
\end{center}
\caption{For the 36 combinations of {\sc Disperse} web and cluster
  matching, the ratio of the maximum of the average filament pair profile
  peaks (for counts) over the average pair profile peak, as a function
  of the fraction of all cluster pairs which are filaments for that
  given combination. The general trend within a matching method is that larger fraction of pairs with filaments have less enhancement of the peak density over that of the average, as is expected. Lengths are denoted by shape, triangles are for maximum length 40 $Mpc/h$, crosses for 50 $Mpc/h$ and circles for 60 $Mpc/h$. Shorter maximum lengths have higher fractions of cluster pairs linked by filaments. Points further towards the top are for smaller smoothing and higher mass subhalos.
}
\label{fig:overdens}
\end{figure}

\subsection{Subhalo counts and halo mass density perpendicular to cluster pairs}
\label{sec:pairdens}
One can look at profiles of
different quantities perpendicular to the axis between cluster pairs, both those with and without filaments, for any web definition. If the cluster pair is
connected by a filament, this is the filamentary radial profile of that
quantity.  Profiles for two
quantities are shown for all
cluster pairs up to some maximum separation, in
Fig.~\ref{fig:pairprof} for the (11996, 20906, and 33842) cluster
pairs with separations less than ($40 Mpc/h$, $50 Mpc/h$, $60 Mpc/h$),
respectively.  At left is the profile in subhalo number counts (in
principle these would be associated with galaxies in a semi-analytic
model), at right are the halo mass densities (where the mass of each halo is located
at the halo center, i.e., the halo profile is neglected).  Different
subhalo infall mass (at left) and halo mass ranges (at right) are separated out, ($10^{11}M_\odot -10^{12}M_\odot$, $10^{12}M_\odot-10^{13}M_\odot$, $10^{13}M_\odot-10^{14}M_\odot$), as
indicated.  The two panels below show the profiles rescaled to their values at 4 $Mpc/h$ and then shifted to separate out the different subhalo (at left) and halo (at right) mass ranges.
Pairs separated by more than 60 $Mpc/h$ are omitted, ranging from
0.3\% to 6\% of the cluster pairs with web-assigned filaments.  As can
be inferred from Fig.~\ref{fig:probfil}, these omitted filament pairs
are a very small fraction of all large separation pairs.  (Larger fractions
of long filaments appeared with larger web smoothings.) The mass
density is highest for the highest mass halos, and smallest for the
pairs with the largest separation (presumably in part because larger pair
separations result in including more pairs with no filaments).  

For cluster pairs with filaments, several profiles corresponding to an
underlying {\sc Disperse} web with 2 $Mpc/h$ smoothing, 3$\sigma$
persistence, matched to clusters via ``patch,'' are shown
in Fig.~\ref{fig:pairfilprof}. The red and blue lines indicate the profiles for cluster pairs assigned filaments 
which interpolate through an unmatched {\sc Disperse} node which are
either kept (``c'' fils) or dropped (``dir c'' fils), the two profiles
are not that different. Also shown is the average profile between all cluster
pairs with the same maximum separation and subhalo minimum mass, and
the average 
profile for the cluster pairs
without a filament (slightly lower than the average profile).  The {\sc
  Disperse} node pair profile, dotted line, is between all {\sc Disperse} node pairs in the corresponding web.
For $r> 2 Mpc/h$ and visually detected filaments, \citet{ColKruCon05}
found a $1/r^2 $ profile, similar fall off was found by \citet{AravanJon10} for a multiscale web finder.   The average counts and mass profiles averaged over cluster filament pairs tended to be
closer to a $1/r$ profile, with the stacked halo mass profile seeming
slightly shallower than the counts profile.
This might indicate some issue with the assigned filaments, either in
how they are chosen, or the fact that filaments are taken to be
straight between cluster pairs, which might be washing out some of
the radial structure.  In the \citet{ColKruCon05} filament catalogue, only 38\% of the filaments
were straight lines between the cluster endpoints.  For the cluster pairs which mapped directly to {\sc Disperse} filaments, with no interpolation, the offset between the line connecting the cluster (or {\sc Disperse} node) endpoints and the filament centers (the saddle critical points provided by {\sc
  Disperse}) had a median value ranging from 1-3 $Mpc/h$ for the different web and matching variations.

As the count profiles tend to have a similar shape,
one can characterize the filament enhancement of counts relative to counts around all cluster pairs with the same separation and subhalo mass ranges at a given distance from the cluster pair axis, chosen here to be 1.5 $Mpc/h$. The
distribution of this filament associated density enhancement is shown as a function of the
fraction of cluster pairs linked by filaments in
Fig.~\ref{fig:overdens}. (As most of the cluster pairs don't have
filaments, the difference between profiles for the unlinked cluster
pairs and all cluster pairs for maximum lengths $\geq 40 Mpc/h$ is much smaller than that between linked and all cluster pairs.)  The fraction of cluster pairs which have filaments goes down
with increasing maximum cluster separation, and decreases as well as with more
restrictive ways of matching clusters to filaments, ranging from
$\sim$ 30\% of the pairs for one instance of ``patch'' and 40 $Mpc/h$
separation, to 2\% of the pairs for ``pix'' and 60
$Mpc/h$ separation, roughly integrals of the distributions shown in
Fig.~\ref{fig:probfil}. 
There tends to be less density enhancement relative to all
pairs as the fraction of pairs which are filaments increases, as is
expected, and the
enhancement also tends to be lower for the' ``pix'' matching and highest
for the ``patch'' matching, as can be seen in Fig.~\ref{fig:overdens}.  The enhancement relative to all pairs is larger for the higher mass tracers, expected from higher mass having higher bias.


On average, cluster pairs with {\sc Disperse} web assigned filaments
were a relatively small fraction of close cluster pairs. Nonetheless,
the average subhalo count density found here, around 1.5 $Mpc/h$ from
the cluster-cluster line and irrespective of whether a filament was
present, was enhanced by a factor of around 3 relative to its value
$\sim$10 $Mpc/h$ away from the line.  Pairs with filaments assigned
for any of the 36 combinations had maximum subhalo counts ranging from
$\sim$5-8 times the subhalo counts $\sim$ 10 $Mpc/h$ away from the center. 

Subhalo count profiles and halo mass profiles were also calculated for
{\sc Disperse} node pairs connected by filaments, one of which is shown in Fig.~\ref{fig:pairfilprof} for the corresponding underlying {\sc Disperse} web.
There are several uncertainties and subtleties in calculating {\sc Disperse} filament profiles.  The {\sc Disperse} node filaments are taken
to run straight between the two nodes, discarding the shape
information provided by {\sc Disperse}.  
In addition, the {\sc Disperse} nodes
are defined for pixel (2 $Mpc/h$ side) averaged densities, so that 
actual density peaks, if present, might be anywhere in the pixel, and are likely offset with respect to the few discrete positions the
{\sc Disperse} nodes take.  In practice, the {\sc Disperse}
filament node pair profiles are stacked just as those for cluster
pairs. 
The subhalo count peak (at 1.5 $Mpc/h$) overdensity between {\sc Disperse} nodes
ranges between 
0.6 and $\sim 2$ of its counterpart for all cluster pairs (going up to the
same maximum separation), with the highest relative values
corresponding to 
the largest maximum separation pairs and longest {\sc
  Disperse} filaments.  For halo mass overdensity,  the ratio ranged
from 0.6 to 2.6 for the peak (again at 1.5 $Mpc/h$) for  {\sc Disperse} node pairs connected by filaments relative to all cluster pairs, again with the samples including the longest filaments and largest cluster separations giving the largest values. This increase was likely in part due to large separation cluster pairs being less likely to have filaments between them. 

\section{Summary and Discussion}
\label{sec:disc}
Here, the largest mass halos in the universe, clusters $10^{14} M_\odot$ and above, were matched to nodes in cosmic webs created using the web finder {\sc Disperse} on pixel overdensities in a
fixed time N-body simulation box.  Filaments between cluster matched {\sc
  Disperse} nodes
were assigned to their corresponding cluster pairs.

 {\sc Disperse} was chosen because of
its public availability and its frequent use, and applied to 4
smoothings of the underlying dark matter, the largest of which (5
$Mpc/h$) matched too poorly to clusters to be usefully studied in much detail.
For the remaining smoothings, 3 values of the {\sc Disperse}
persistence parameter were considered, resulting in 9 different {\sc
  Disperse} webs based upon the same underlying smoothed simulation.  

In each of these 9 webs, clusters were matched to nodes
by assigning a volume to the {\sc Disperse} nodes via 4 methods, and then seeing if
the cluster lay within this volume.  Volumes were given by the pixel
of the {\sc Disperse} node
($\sim$ 2 $Mpc/h$), or a twice smoothing radius sphere around the {\sc Disperse} node, or within the
same Hessian node ``patch'' (continguous pixels all classified as
nodes via the Hessian method).  No method matched every cluster to a
{\sc Disperse} node, but about half of the clusters had a matched {\sc Disperse} node for
every method, in all 9 underlying {\sc Disperse} webs, and 3/4 had matches if
the most restrictive matching (lying in the same pixel) was dropped.
The clusters matched to {\sc Disperse} nodes for the most methods and
underlying {\sc Disperse} webs tended to have
higher mass and lower ``velocity shear'', and, perhaps a higher likelihood of recent 1:3 mergers.  High density {\sc Disperse} nodes were more
likely to have a matched cluster.

 For every {\sc Disperse} web, there generally appeared to be a distinct population of
clusters ``near'' the {\sc Disperse} nodes, and for clusters and their
nearest {\sc Disperse}
nodes within twice the smoothing length of each other, a cluster mass-{\sc
  Disperse} node density relation is seen.

{\sc Disperse} filaments where both {\sc Disperse} node endpoints have a matched cluster can be assigned
to the corresponding cluster pair; a method for interpolating filaments through
unmatched {\sc Disperse} nodes was also applied.  For smoothings $<$ 5 $Mpc/h$ and the ``nearest/fixed/patch'' methods, 10\%
-25\% of clusters have no filament, either because they
 had no matching {\sc Disperse} node (and thus did not get assigned a  {\sc
   Disperse} node's filaments)
 or because their cluster matched {\sc
   Disperse} nodes were not linked to any other cluster matched {\sc
   Disperse} node.  Only 47 of the clusters (out of 2898) never 
 had a filament for any of the underlying {\sc Disperse} webs and matching
 methods (152 clusters if one didn't interpolate through unmatched {\sc Disperse nodes} to assign filaments).  Closer cluster pairs
 were more likely to be assigned filaments, although some of the
 matching methods showed a dip in this probability between 5-10
 $Mpc/h$. Beyond 15-20 $Mpc/h$ separations, pairs with filaments were
 a minority of cluster pairs (but around half of the cluster filament pairs
 were at larger separations).  
 Cluster pairs with the long
 axis of one cluster aligned with the direction of the cluster pair also made
 the cluster pair more likely to be linked by a filament, and 
 ``importance'' from an out of the box machine learning method also
 found that cluster-cluster long axis alignments made an intervening
 filament more likely, as expected from other studies.  The average subhalo counts and mass weighted halo counts around cluster pairs are enhanced, with more
 enhancement for the subset of cluster pairs connected by filaments.  The profile
 away from the filament pair was closer to $1/r$, weaker than the $1/r^2$
 falloff beyond 2 $Mpc/h$, as found in e.g. \citet{ColKruCon05,AravanJon10},
 perhaps due to assuming here that the filaments were straight lines between the
 pairs, but also possibly due to how filaments are
 defined from the underlying web.  Unlike the cluster-{\sc Disperse}
 node matching, where many clusters had {\sc Disperse}
 node matches for most of the
 {\sc Disperse} web variations and matching methods, most
 cluster pairs only had filaments for a few of the
 underlying {\sc Disperse} web variations and matching methods.

This matching between clusters and the cosmic web picks out
subclasses of both, that is, clusters and cluster pairs which have 
corresponding {\sc Disperse} nodes and filaments, and
vice versa.
Some trends in unmatched clusters were seen, and it would be
interesting to check whether others are present
 either in the dark matter properties or
observable properties of these clusters. Unfortunately, the more recent
galaxy formation models built upon the Millennium simulation are only for the
rescaled \citep{AngWhi10,AngHil15} Millennium simulation, for which smooth particle densities
were not available.  It would be interesting to study the different
kinds of clusters and cluster pairs which arose here in a simulation incorporating what is now known about current galaxy observables.

The unmatched {\sc Disperse} nodes tended to be less massive but not
always; those in the higher mass range, where sometimes there is a
cluster match and sometimes there isn't, would also be interesting to
better understand.  For example, nodes also can be evaluated in terms of other
properties such as their histories \citep{Cad20}, just as clusters are. It also is possible to go down to lower halo (or subhalo) mass and do the
matching procedure again.  With more {\sc Disperse} nodes matched to halos, the resulting network of halos (including clusters) and their inherited filaments will include more of the underlying {\sc Disperse} web.

The clusters also give a reference point with which to compare nodes
between different webs.  That is, as 3/4 of the clusters
had {\sc Disperse} node matches for
``nearest'',''fixed'', and ``patch'' matching methods and all 9 webs, the {\sc
  Disperse} nodes they match to can be intercompared.  This could
also be done for other webs, as a way to identify corresponding nodes.

In this approach, cluster-cluster filament pairs are a particular selection of all the
cluster pairs, and different underlying {\sc Disperse} webs have
different cluster pairs being assigned filaments.  Unlike
cluster-{\sc Disperse} node matching, where many clusters always had
matches, fewer than 10\% of
the cluster pairs which ever had an assigned filament had a filament for
most of the different webs and matching methods; this
fraction dropped by a factor of 2 if interpolation through dropped
{\sc Disperse} nodes was dropped.  
It would be interesting to see how much overlap there is with filaments found
between clusters for other constructions which also have some sort of
node-cluster identification (or have clusters defined as nodes directly).

At least for the methods of assigning filaments to cluster pairs here, 
whether a cluster pair shares a
filament depends upon which web definition is of interest, which is a
question which comes from outside of the cluster-{\sc Disperse} node correspondence, and of course, other web finders could be used as well.  One might use some of these differences between
cluster filament pair properties (denser
filament profile, or higher likelihood of nearby pairs being
connected) as ways to characterize, and maybe select, web
finders, depending upon the application in mind.

Clusters are very important gravitationally bound objects in the
universe, and special environments in galaxy formation, while
nodes anchor the cosmic web which evolves as structure forms.  The connection between the two provides a way to better understand clusters, cosmic web nodes, and their relation and evolution together.

\section*{Acknowledgements}
Many thanks especially to K. Kraljic and M. White for numerous discussions and suggestions, and to M. Alpaslan, S. Codis, C. Laigle, C. Pichon, and A. White for help as well, and to the participants of the Higgs 2019 Cosmic Web meeting, and the CCA, CERN, IAP, NYU, and the Royal Observatory of Edinburgh for hospitality and opportunities to present and discuss this work. I am also grateful to the referee, C. Miller, for many helpful questions, criticisms and suggestions, which resulted in my making many improvements.
And I am also grateful for the Millennium simulation database.

\section*{Data availability}
The simulation data was downloaded from the Millennium simulation database using queries which are provided in the footnotes.  The cosmic webs were constructed via {\sc Disperse}, by using the package available at {\tt http://www2.iap.fr/users/sousbie/web/html/indexd41d.html } and then running the commands described in footnotes in the text.

\end{document}